\documentclass[aps,twocolumn]{revtex4}

\usepackage{graphics}
\usepackage{epsfig}
\usepackage{bm}
\usepackage{amsmath,amssymb}

\usepackage{graphics}
\usepackage{epsfig}

\usepackage{amsmath}
\usepackage{amssymb}
\usepackage{stmaryrd}

\def\ov#1{\overline{#1}}

\def\pd#1#2{\frac{\partial #1}{\partial #2}}

\newcommand{\bc}{\begin{center}}
\newcommand{\ec}{\end{center}}
\newcommand{\bt}{\begin{tabbing}}
\newcommand{\et}{\end{tabbing}} 
\newcommand{\be}{\begin{eqnarray*}}
\newcommand{\ee}{\end{eqnarray*}}
\newcommand{\bs}{\begin{slide}}
\newcommand{\es}{\end{slide}}

\begin{document}

\title{Asymptotic Limit-cycle Analysis of Oscillating Chemical Reactions}

\author{Alain J.~Brizard$^{1}$ and Samuel M.~Berry$^{2}$}
\affiliation{$^{1}$Department of Physics and $^{2}$Department of Chemistry,  Saint Michael's College, Colchester, VT 05439, USA} 

\begin{abstract}
The asymptotic limit-cycle analysis of mathematical models for oscillating chemical reactions is presented. In this work, after a brief presentation of mathematical preliminaries applied to the biased Van der Pol oscillator, we consider a two-dimensional model of the Chlorine dioxide Iodine Malonic-Acid (CIMA) reactions and the three-dimensional and two-dimensional Oregonator models of the Belousov-Zhabotinsky (BZ) reactions. Explicit analytical expressions are given for the relaxation-oscillation periods of these chemical reactions that are accurate within 5\% of their numerical values. In the two-dimensional CIMA and Oregonator models, we also derive critical parameter values leading to canard explosions and implosions in their associated limit cycles.
\end{abstract}

\date{\today}


\maketitle

\section{Introduction}

Oscillating chemical reactions have attracted attention since their first announcement \cite{Noyes_Field_1974,Zhabotinsky_1991}. Observed as early as the 17th century, it was not until the early 20th century that they came into a modern framework with the “iron nerve” and “mercury heart” reactions (see Refs. \cite{Zhabotinsky_1991,Cervellati_Greco_2017} for historical surveys). The newly discovered oscillatory reactions, however, posed a problem as they seemed to defy the second law of thermodynamics. Thus, it was not until the mid-1960s that the proposed mechanisms for oscillatory reactions would become accepted. 

In 1951, Belousov found one of the most famous examples of an oscillatory system, the BZ reaction (see Ref. \cite{Zhabotinsky_1991} and references therein) named after him and Zhabotinsky who furthered Belousov’s research. Belousov’s results were first published in 1959 and laid the groundwork for the future emergence of the field’s study. By 1972, increased interest in chemical oscillators came from papers puiblished by Field {\it et al.} \cite{FKN_1972} and Winfree \cite{Winfree_1972}, among others, detailing a more complete mechanism of the BZ reaction and chemical reaction-diffusion systems, respectively. Further research by Clarke \cite{Clarke_1974} paved the way for steady-state stability analysis.

In addition to the BZ reaction, numerous other chemical oscillators have been found, such as the chemical clock reaction used in chemistry demonstrations. Various other types of oscillating systems outside of chemical reactions have also been found (see Ref \cite{Noyes_Field_1974} for an overview of the various systems as well as a mathematical overview of the subject matter). The purpose of the present paper is to perform a unified asymptotic analysis of two well-known oscillating chemical reactions:  The Chlorine dioxide Iodine Malonic-Acid (CIMA) reaction and the Oregonator model of the Belousov-Zhabotinsky (BZ) reactions.

The remainder of the paper is organized as follows. In Sec.~\ref{sec:math}, we present the mathematical preliminaries associated with the type of coupled first-order differential equations considered in our work. In particular, we present the asymptotic analysis leading to an explicit integral expression for the period of large-amplitude relaxation oscillations. We also present the {\it canard}-behavior analysis that predicts the sudden appearance (and possible disappearance) of these large-amplitude relaxation oscillations from small amplitude periodic oscillations. 

Next, in Sec.~\ref{sec:vdp}, we apply these mathematical preliminaries to the relaxation oscillations associated with the biased Van der Pol model, where we show that the period and canard behavior of these relaxation oscillations are accurately predicted by the asymptotic formulas derived in Sec.~\ref{sec:math}.

In the next two Sections, we focus our attention on two well-known paradigm models for oscillatory chemical reactions: the Chlorine dioxide Iodine Malonic-Acid (CIMA) reactions (Sec.~\ref{sec:CIMA}) and the Oregonator model of the Belousov-Zhabotinsky (BZ) reactions (Sec.~\ref{sec:Oregonator}), presented first as a three-variable model (Oregonator-3) and then reduced to a two-variable model (Oregonator-2). Once again, we show that our asymptotic formulas for the period of relaxation oscillations as well as their canard appearance (explosion) and disappearance (implosion) can be accurately predicted. Much of the success of these formulas is credited to our ability of finding analytical expressions for the roots of cubic polynomials with coefficients that are functions of model parameters (the general method is presented in App.~\ref{sec:cubic}).

\section{\label{sec:math}Mathematical Preliminaries}

In the present paper, we transform two-variable chemical kinetic equations into dimensionless nonlinear first-order ordinary differential equations, which are generically expressed as
\begin{equation}
\left. \begin{array}{rcl}
\dot{x} &=& F(x,y; a) \\
\dot{y} &=& \epsilon\,G(x,y; a)
\end{array} \right\},
\label{eq:xy_general}
\end{equation}
where $x$ and $y$ denote dimensionless chemical concentrations, and each dimensionless time derivative is represented with a dot (e.g., $\dot{x} = dx/dt$). On the right side of Eq.~\eqref{eq:xy_general}, the dimensionless parameter $\epsilon$ plays an important role in the qualitative solutions of Eq.~\eqref{eq:xy_general}, while the functions $F(x,y;a)$ and $G(x,y;a)$ (which may depend on a dimensionless parameter $a$) are used to define the nullcline equations: $F(x,y;a) = 0 = G(x,y;a)$, which yield separate curves $y = f(x;a)$ and $y = g(x;a)$ onto the $(x,y)$-plane. A simplifying assumption used in the models investigated in this paper is that the functions $F$ and $G$ are at most separately linear in $y$ and $a$, with $\partial^{2}F/\partial y\partial a = 0 = \partial^{2}G/\partial y\partial a$.

By rescaling the dimensionless time $\tau \equiv \epsilon\,t$, Eq.~\eqref{eq:xy_general} is transformed into a new set of first-order differential equations
\begin{equation}
\left. \begin{array}{rcl}
\epsilon\,x^{\prime} &=& F(x,y; a) \\
y^{\prime} &=& G(x,y; a)
\end{array} \right\},
\label{eq:xy_slow}
\end{equation}
where a prime now denotes a derivative with respect to $\tau$. Since $\epsilon \ll 1$ in our analysis, the variables $x$ and $y$ are known as the fast and slow variables, respectively. Because $\epsilon$ appears on the left side of Eq.~\eqref{eq:xy_slow}, these equations are known as singularly perturbed equations.

We note that the slope function $m(x,y;a) \equiv \dot{y}/\dot{x} = y^{\prime}/x^{\prime}= \epsilon\,G(x,y;a)/F(x,y;a)$ is a useful qualitative tool as we follow an orbit in the $y(t)$-versus-$x(t)$ phase space. In particular, we see that the orbit crosses the $y$-nullcline horizontally $(m = 0)$ and it crosses the $x$-nullcline vertically $(m = \pm\infty)$. Hence, in the limit $\epsilon \ll 1$, the slope function is near zero (i.e., the orbit is horizontal) unless the orbit is near the $x$-nullcline, where $F(x,y;a) \simeq 0$. As the slope $m(x,y;a)$ depends on the model parameter $a$, the shape of the orbit solution will also change with $a$.

\subsection{Linear stability analysis}

If these nullcline curves intersect at $(x_{0},y_{0})$, where $x_{0} = x_{0}(a)$ and $y_{0}(a) = f(x_{0}) = g(x_{0})$, the point $(x_{0},y_{0})$ is called a fixed point of Eq.~\eqref{eq:xy_general}. The stability of this fixed point is investigated through a standard normal-mode analysis \cite{Strogatz_2015}, where $x = x_{0} + \delta\ov{x}\,\exp(\lambda t)$ and $y = y_{0} + \delta\ov{y}\,\exp(\lambda t)$ are inserted into Eq.~\eqref{eq:xy_general} to obtain the linearized matrix equation
\begin{equation}
\left( \begin{array}{cc}
\lambda - F_{x0} & -\,F_{y0} \\
-\,\epsilon\,G_{x0} & \lambda - \epsilon\,G_{y0}
\end{array} \right) \cdot \left(\begin{array}{c} 
\delta\ov{x} \\
\delta\ov{y}
\end{array} \right) \;=\; 0,
\label{eq:Jac}
\end{equation}
where the constant eigenvector components $(\delta\ov{x},\delta\ov{y})$ are non-vanishing only if the determinant of the linearized matrix vanishes. Here, $(F_{x0},F_{y0})$ and $(G_{x0},G_{y0})$ are partial derivatives evaluated at the fixed point $(x_{0},y_{0})$ and the eigenvalues $\lambda_{\pm} = \frac{1}{2}\;\tau \pm \frac{1}{2}\, \sqrt{\tau^{2} - 4\;\Delta}$ are roots of the quadratic characteristic equation $\lambda^{2} - \tau\,\lambda + \Delta = 0$, where $\tau(a,\epsilon) \equiv F_{x0} + \epsilon\,G_{y0} = \lambda_{+} + \lambda_{-}$ and $\Delta(a,\epsilon) = \epsilon\,(F_{x0}\,G_{y0} - F_{y0}\,G_{x0}) = \lambda_{+}\cdot \lambda_{-}$ are the trace and determinant of the Jacobian matrix, respectively.

The fixed point is a stable point ($\tau < 0$ and $\Delta > 0$) that is either a node $(\tau^{2} > 4\;\Delta)$, when the eigenvalues are real and negative: $\lambda_{-} < \lambda_{+} < 0$, or a focus $(\tau^{2} < 4\;\Delta)$, when the eigenvalues are complex-valued ( $\lambda_{-} = \lambda_{+}^{*}$) with a negative real part. Otherwise, the fixed point is either an unstable point ($\tau > 0$ and $\Delta > 0$) or a saddle point $(\Delta < 0)$. 

Periodic solutions of Eq.~\eqref{eq:xy_general} exist when a Hopf bifurcation \cite{Strogatz_2015} replaces an unstable fixed point with a stable limit cycle, which forms a closed curve in the $(x,y)$-plane. Here, a limit cycle appears when the $x$-nullcline function $f(x;a)$ has non-degenerate minimum and maximum points and it is stable whenever the trace $\tau(a) > 0$ is positive in the range $a_{s} < a < a_{u}$.

\subsection{Asymptotic limit-cycle period}

We shall see that, in the asymptotic limit $\epsilon \ll 1$, the limit-cycle curve is composed of segments that are close to the $x$-nullcline. In this limit, the asymptotic period can be calculated as follows. First, we begin with the $x$-nullcline $y = f(x;a)$ on which we obtain $dy/dt = f^{\prime}(x;a)\,dx/dt$. Next, we use the $y$-equation $dy/dt = \epsilon\,G(x,y;a)$, into which we substitute the $x$-nullcline equation: $dy/dt = \epsilon\,G(x,\,f(x;a);a)$. 

By combining these equations, we obtain the infinitesimal asymptotic-period equation $\epsilon\,dt = f^{\prime}(x;a)\,dx/G\left(x, f(x;a);a\right)$, which yields the asymptotic limit-cycle period
\begin{eqnarray}
\epsilon\,T_{\rm ABCDA}(a) &=& \int_{x_{A}(a)}^{x_{B}(a)} \frac{f^{\prime}(x;a)\;dx}{G(x,f(x;a);a)} \nonumber \\
 &&+\; \int_{x_{C}(a)}^{x_{D}(a)} \frac{f^{\prime}(x;a)\;dx}{G(x,f(x;a);a)}.
\label{eq:vdP_period}
\end{eqnarray}
Here, the asymptotic limit cycle ABCDA combines the slow $x$-nullcline orbits $x_{A} \rightarrow x_{B}$ and $x_{C} \rightarrow x_{D}$ and the fast horizontal transitions $x_{B} \rightarrow x_{C}$ and $x_{D} \rightarrow x_{A}$, which are ignored in Eq.~\eqref{eq:vdP_period}. Generically, the values $x_{D}(a) < x_{B}(a)$ are the minimum and maximum of the $x$-nullcline $y = f(x;a)$, respectively, where $f^{\prime}(x;a)$ vanishes. The points $x_{C}(a) < x_{A}(a)$, on the other hand, are the minimum and maximum of the asymptotic limit cycle.

\subsection{Canard transition to relaxation oscillations}

Whenever the fixed point $x_{0}(a)$ comes close to a critical point of the $x$-nullcline, either $x_{B}(a)$ or $x_{D}(a)$, a transition involving a bifurcation to a large-amplitude relaxation oscillation becomes possible. This transition, which occurs suddenly as the model parameter $a$ crosses a critical value $a_{c}(\epsilon)$, is referred to as a {\it canard} explosion or implosion, depending on whether the large-amplitude relaxation oscillation appears or disappears. For a brief review of the early literature on canard explosions, see Refs.~\cite{Diener_1984,Wechselberger_2007} and references therein. For a mathematical treatment, on the other hand, see Refs.~\cite{Krupa_2001,Fenichel_1979}.

We now present a perturbative calculation of the critical canard parameter $a_{c}(\epsilon)$ as an asymptotic expansion in terms of the small parameter $\epsilon$. For this purpose, we use the invariant-manifold solution $y = \Phi(x,\epsilon)$ of geometric singular perturbation theory \cite{Fenichel_1979,Ginoux_2011}, which yields the generic canard perturbation equation
\begin{eqnarray}
\dot{y} &=& \epsilon\,G\left(x, \Phi(x,\epsilon);\frac{}{}  a\right) \;=\; \pd{\Phi(x,\epsilon)}{x}\;\dot{x} \nonumber \\
 &=& \pd{\Phi(x,\epsilon)}{x}\;F\left(x, \Phi(x,\epsilon);\frac{}{} a\right),
 \label{eq:canard_eq}
\end{eqnarray}
where $\Phi(x,\epsilon) = \sum_{k=0}^{\infty}\epsilon^{k}\Phi_{k}(x)$ and $a_{c}(\epsilon) = \sum_{k=0}^{\infty}\epsilon^{k}a_{k}$. At the lowest order $(\epsilon = 0)$, we find
\begin{equation}
0 \;=\; F\left(x, \Phi_{0}(x);\frac{}{}a_{0}\right),
\label{eq:canard_eq_0}
\end{equation}
which yields the lowest-order $x$-nullcline 
\begin{equation}
\Phi_{0}(x) \;\equiv\; f(x;a_{0}). 
\end{equation}

\subsubsection{First-order perturbation analysis}

At the first order in $\epsilon$, we now find from Eq.~\eqref{eq:canard_eq}:
\begin{equation}
G(x,\Phi_{0};a_{0}) \;=\; \Phi_{0}^{\prime}(x) \left[ F_{y0}\;\Phi_{1}(x) \;+\frac{}{} F_{a0}\;a_{1} \right],
\label{eq:canard_1_eq}
\end{equation}
where $F_{y0} = (\partial F/\partial y)_{0}$ and $F_{a0} = (\partial F/\partial a)_{0}$ are evaluated at $(x, \Phi_{0}; a_{0})$, and $\Phi_{0}^{\prime}(x)$ can be factored as
\begin{equation}
\Phi_{0}^{\prime}(x) \;\equiv\; (x - x_{c})\,\Psi_{0}(x),
\label{eq:Phi0_prime}
\end{equation}
where $\Psi_{0}(x)$ is assumed to be finite at the critical point $x = x_{c}(a_{0})$ (i.e., a minimum or a maximum of the $x$-nullcline). Since the right side of Eq.~\eqref{eq:canard_1_eq} vanishes at the critical point $x_{c}(a_{0})$, we find $G(x_{c},\Phi_{0c};a_{0}) = 0$, which implies the identity
\begin{equation}
x_{0}(a_{0}) \;\equiv\; x_{c}(a_{0}).
\label{eq:x_0c}
\end{equation}
Hence, the fixed point $x_{0}$ has merged with the critical point $x_{c}$ of the $x$-nullcline at a unique value $a_{0}$, i.e., the fixed point $x_{0}(a_{0})$ is either at the maximum $x_{0}(a_{0}) = x_{B}(a_{0})$, which yields $a_{0} = a_{B0}$, or at the minimum $x_{0}(a_{0}) = x_{D}(a_{0})$, which yields $a_{0} = a_{D0}$. 

With this choice of $a_{0}$, we can write the factorization
\begin{equation}
G(x,\Phi_{0};a_{0}) \;\equiv\; (x - x_{c})\;H_{1}(x),
\label{eq:H1_factor}
\end{equation}
where $H_{1}(x)$ is finite at $x = x_{c}(a_{0})$. Hence, from Eq.~\eqref{eq:canard_1_eq}, we obtain the first-order solution
\begin{equation}
\Phi_{1}(x) \;\equiv\; K_{1}(x) \;-\; h(x)\;a_{1},
\label{eq:canard_1_sol}
\end{equation}
where we introduced the definitions
\begin{equation}
\left. \begin{array}{rcl}
K_{1}(x) &\equiv& H_{1}(x)/[\Psi_{0}(x)\,F_{y0}(x)] \\
 && \\
 h(x) &\equiv& F_{a0}(x)/F_{y0}(x)
 \end{array} \right\},
 \label{eq:canard_eq_1}
 \end{equation}
 which are both finite at $x = x_{c}(a_{0})$.

\subsubsection{Second-order perturbation analysis}

At the second order in $\epsilon$, we find from Eq.~\eqref{eq:canard_eq}:
\begin{eqnarray}
G_{y0}\;\Phi_{1} \;+\; G_{a0}\;a_{1} &=& \Phi_{0}^{\prime} \left( F_{y0}\;\Phi_{2} \;+\frac{}{} F_{a0}\;a_{2} \right) \nonumber \\
 &&+\; \Phi_{1}^{\prime} \left( F_{y0}\;\Phi_{1} \;+\frac{}{} F_{a0}\;a_{1} \right) \nonumber \\
  &=& \Phi_{0}^{\prime} F_{y0}\left( \Phi_{2} \;+\frac{}{} h\;a_{2} \right) \nonumber \\
 &&+\; F_{y0}\,\left( K_{1}^{\prime} \;-\frac{}{} h^{\prime}\,a_{1}\right) K_{1},
\label{eq:canard_eq_2}
 \end{eqnarray}
 where $G_{y0} = (\partial G/\partial y)_{0}$ and $G_{a0} = (\partial G/\partial a)_{0}$ are evaluated at $(x, \Phi_{0}; a_{0})$, and we have used the first-order solution \eqref{eq:canard_1_sol}. By rearranging terms in Eq.~\eqref{eq:canard_eq_2}, we obtain the second-order equation
\begin{equation}
S_{1}(x)\; a_{1} \;-\; R_{2}(x) \;=\; \Phi_{0}^{\prime}(x) \left[F_{y0}\;\Phi_{2}(x) \;+\frac{}{} F_{a0}\;a_{2} \right],
\label{eq:canard_2}
\end{equation}
where we introduced the definitions
\begin{eqnarray}
R_{2}(x) & = & K_{1}(x) \left[ F_{y0}\; K_{1}^{\prime}(x) \;-\frac{}{} G_{y0}\right], 
\label{eq:canard_R2} \\
S_{1}(x) & = &  G_{a0} -  G_{y0}\;h(x) + F_{y0}\,h^{\prime}(x)\,K_{1}(x),
\label{eq:canard_S1}
\end{eqnarray}
which are both finite at $x_{c}(a_{0})$.  

Once again, since the right side of this equation vanishes at the critical point $x = x_{c}(a_{0})$, the left side must also vanish, and we obtain the first-order correction 
\begin{equation}
a_{1} \;=\; R_{2}(x_{c})/S_{1}(x_{c}).
\label{eq:canard_a1}
\end{equation}
By factoring the left side of Eq.~\eqref{eq:canard_2},
\begin{equation}
S_{1}(x)\; a_{1} \;-\; R_{2}(x) \;=\; (x - x_{c})\;H_{2}(x),
\label{eq:H2_factor}
\end{equation}
we now obtain the second-order solution
\begin{equation}
\Phi_{2}(x) \;\equiv\; K_{2}(x) \;-\; h(x)\,a_{2},
\label{eq:canard_2_sol}
\end{equation}
where $K_{2}(x) \equiv H_{2}(x)/[\Psi_{0}(x)F_{y0}(x)]$ and $h(x)$ is defined in Eq.~\eqref{eq:canard_eq_1}.

\subsubsection{Higher-order perturbation analysis}

By continuing the perturbation analysis at higher order $(n \geq 3)$, Eq.~\eqref{eq:canard_eq} yields the $n$th-order equation 
\begin{equation}
S_{1}(x)\,a_{n-1} - R_{n}(x) = \Phi_{0}^{\prime}(x)\,F_{y0}\,\left[\Phi_{n}(x) + h(x)\,a_{n}\right],
\label{eq:canard_n}
\end{equation}
where $S_{1}(x)$ is defined in Eq.~\eqref{eq:canard_S1} and
\begin{eqnarray}
R_{n}(x) & = & K_{1}(x)\,F_{y0}\,K_{n-1}^{\prime}(x) \;-\; G_{y0}\,K_{n-1}(x) \nonumber \\
 &&+ \sum_{k=1}^{n-2}F_{y0}\left[K_{k}^{\prime}(x) - h^{\prime}(x)\,a_{k}\right]K_{n-k}(x).
 \label{eq:Rn_sol}
 \end{eqnarray}
 Hence, the left side of Eq.~\eqref{eq:canard_n} vanishes at $x_{c}$ if 
 \begin{equation}
 a_{n-1} \;=\; R_{n}(x_{c})/S_{1}(x_{c}), 
 \label{eq:canard_an-1}
 \end{equation}
 and the $n$th-order solution is obtained by first obtaining the factorization 
 \begin{equation}
 S_{1}(x)\, a_{n-1} \;-\; R_{n}(x) \;=\; (x - x_{c})\,H_{n}(x), 
 \label{eq:Hn_factor}
 \end{equation}
 so that  
 \begin{equation}
\Phi_{n}(x) \;\equiv\; K_{n}(x) \;-\; h(x)\,a_{n},
\label{eq:canard_n_sol}
\end{equation}
where  $K_{n}(x) \equiv H_{n}(x)/[\Psi_{0}(x)F_{y0}(x)]$ and 
 \begin{equation}
 a_{n} \;=\; R_{n+1}(x_{c})/S_{1}(x_{c}), 
 \label{eq:canard_an}
 \end{equation} 
 is calculated from Eq.~\eqref{eq:Rn_sol}. We note that, once the function $R_{n}(x)$ is calculated in Eq.~\eqref{eq:Rn_sol}, the most computationally-intensive step is the factorization \eqref{eq:Hn_factor}, with $a_{n-1}$ is calculated from Eq.~\eqref{eq:canard_an-1}.

\subsubsection{Critical canard parameter}

As a result of the perturbative solution of Eq.~\eqref{eq:canard_eq}, we have, therefore, calculated the perturbation expansion of the canard critical parameter
\begin{equation}
a_{c}(\epsilon) \;=\; a_{0} \;+\; \frac{1}{S_{1}(x_{c})}\sum_{k=1}^{\infty}\epsilon^{k}\;R_{k+1}(x_{c}).
\label{eq:canard_01}
\end{equation}
For most applications, however, Eq.~\eqref{eq:canard_01} can be truncated at first order in the asymptotic limit $\epsilon \ll 1$: $a_{c}(\epsilon) \simeq a_{0} + a_{1}\,\epsilon$, where $a_{1} > 0$ for a canard explosion, while $a_{1} < 0$ for a canard implosion. 

\section{\label{sec:vdp}Van der Pol Model}

\begin{figure}
\epsfysize=1.8in
\epsfbox{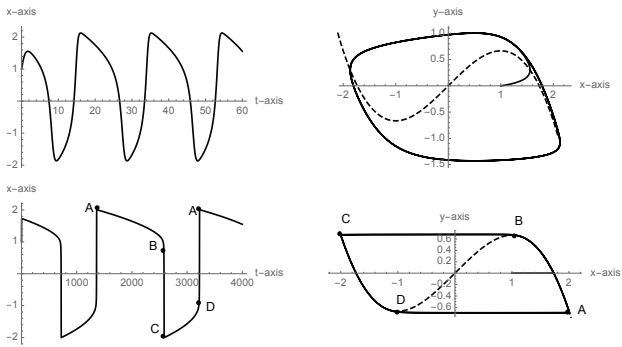}
\caption{Van der Pol solutions $x(t)$ versus time $\tau$ (left column) and phase space plot $y(t)$ versus $x(t)$ (right column), for $a = 0.5$, with the initial condition $x(0) = 1$ and $y(0) = 0$, and $\epsilon = 0.2$ (top row) or $\epsilon = 0.001$ (bottom row). We note that, in the limit $\epsilon \ll 1$, the phase-space orbit has slow segments $A \rightarrow B$ and $C \rightarrow D$ on the $x$-nullcline (shown as a dashed curve) and fast transitions $B \rightarrow C$ and $D \rightarrow A$.}
\label{fig:vdp_x_phase}
\end{figure}

The paradigm model used in our asymptotic analysis is represented by the biased Van der Pol equation \cite{Diener_1984}
\begin{equation}
\ddot{x} \;-\; \nu\,\left(1 - x^{2}\right)\,\dot{x} \;+\; \omega^{2}\,x \;=\; \omega^{2}\,a,
\label{eq:vdp_eq}
\end{equation}
where $\omega$ is the natural frequency of the linearized harmonic oscillator and $\nu$ is the negative dissipative rate, while the bias parameter $a$ represents an equilibrium value of the dimensionless oscillator displacement $x$. 

From Eq.~\eqref{eq:vdp_eq}, we obtain the coupled dimensionless equations
\begin{equation}
\left. \begin{array}{rcl}
\dot{x} & = & x \;-\; x^{3}/3 \;-\; y \\
\dot{y} & = & \epsilon\,(x - a)
\end{array} \right\},
\label{eq:vdp_a}
\end{equation}
where the dimensionless time is normalized to $\nu^{-1}$ and $\epsilon \equiv \omega^{2}/\nu^{2}$ \cite{Footnote_VdP}. Here, the $x$-nullcline is $y(x) = x - x^{3}/3 \equiv \varphi(x)$ (which has a minimum at $x = -1$ and a maximum at $x = 1$) while the $y$-nullcline is a vertical line at $x = a$. The fixed-point is $(x_{0},y_{0}) = (a, a - a^{3}/3)$ and the trace and determinant are $\tau = 1 - a^{2}$ and $\Delta = \epsilon > 0$, respectively. 

It is clear that the fixed-point is stable $(\tau < 0)$ in the range $a^{2} > 1$, while a stable limit cycle exists in the range $-\,1 < a < 1$ (i.e., when the fixed point is located between the minimum and maximum of the $x$-nullcline). Figure \ref{fig:vdp_x_phase} shows the numerical solutions of the Van der Pol equations \eqref{eq:vdp_a} for the case $a = 0.5$ and $\epsilon = 0.2$ (top row) or $\epsilon = 0.001$ (bottom row). Note that, as qualitatively predicted, the orbit crosses the $x$-nullcline vertically (see top right plot). In addition, if $\epsilon$ is small enough, the amplitude $X(\epsilon,a)$ of the Van der Pol oscillation may be approximated as $X(a,\epsilon) \simeq 2$, so that we may take $x_{A} = 2$ and $x_{C} = -2$, with $x_{B} = 1$ and $x_{D} = -1$, as the vertex points of the asymptotic limit cycle $ABCDA$.

\subsection{Asymptotic Van der Pol period}

\begin{figure}
\epsfysize=1.5in
\epsfbox{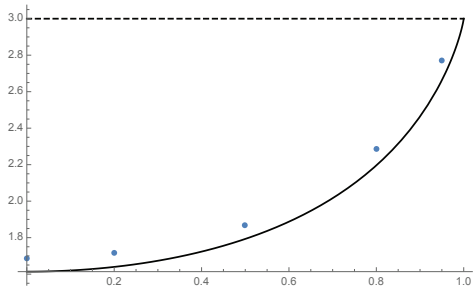}
\caption{Plot of the asymptotic Van der Pol period $\epsilon\,T_{\rm VdP}(a)$ versus the bias parameter $a$, in the limit $\epsilon = 0.001 \ll 1$. The numerical periods $\epsilon\,T_{\rm num}(a)$, shown as dots, are approximately 4\% higher than the asymptotic Van der Pol period \eqref{eq:VdP_period_a}.}
\label{fig:VdP_a}
\end{figure}

In the limit $\epsilon \ll 1$,  the phase-space orbit has slow segments $A\,(x_{A} = 2) \rightarrow B\,(x_{B} = 1)$ and $C\,(x_{C} = -2) \rightarrow D\,(x_{D} = -1)$ on the $x$-nullcline (shown as a dashed curve) and fast horizontal transitions $B\,(x_{B} = 1) \rightarrow C\,(x_{C} = -2)$ and $D\,(x_{D} = -1) \rightarrow A\,(x_{A} = 2)$. The asymptotic period \eqref{eq:vdP_period} for the Van der Pol limit-cycle ABCDA is calculated as
\begin{eqnarray}
\epsilon\,T_{\rm VdP}(a) &=& \int_{2}^{1}\frac{(1 - x^{2})\,dx}{x - a} \;+\; \int_{-2}^{-1}\frac{(1 - x^{2})\,dx}{x - a} \nonumber \\
 &=& 3 \;-\; (1 - a^{2})\;\ln\left(\frac{4 - a^{2}}{1 - a^{2}}\right),
\label{eq:VdP_period_a}
\end{eqnarray}
which is shown in Fig.~\ref{fig:VdP_a}. We note that the asymptotic Van der Pol period \eqref{eq:VdP_period_a} is symmetric in $a$, i.e., $T_{\rm VdP}(-a) = T_{\rm VdP}(a)$. 

The numerical periods $\epsilon\,T_{\rm num}(a)$, which are shown in Fig.~\ref{fig:VdP_a} as dots, are within 4\% higher than the asymptotic Van der Pol period \eqref{eq:VdP_period_a}. These numerical results show that the asymptotic limit $\epsilon \ll 1$ enable us to evaluate the limit-cycle period according to Eq.~\eqref{eq:vdP_period} with excellent accuracy, on both qualitative and quantitative basis. Lastly, we note that the numerical periods are systematically higher than the asymptotic period \eqref{eq:VdP_period_a} because this integral omits the fast transitions $B \rightarrow C$ and $D \rightarrow A$.

\subsection{Canard behavior in the Van der Pol model}

\begin{figure}
\epsfysize=3in
\epsfbox{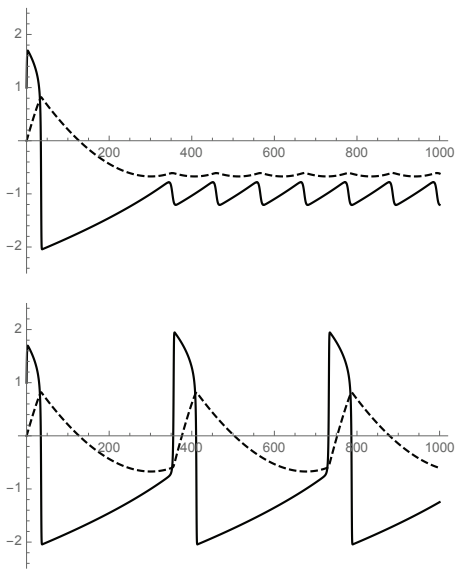}
\caption{Canard explosion in the Van der Pol limit cycle as the bias parameter $a$ crosses the critical value $a_{-}(\epsilon) > -1$ toward the stable limit-cycle regime $(-1 < a < 1)$. Periodic Van der Pol solutions $x(t)$ (solid) and $y(t)$ (dashed) for $\epsilon = 0.01$ and $a =  -\,0.998740$ (top) and $a =  -\,0.998739$ (bottom).}
\label{fig:vdp_minus_canard}
\end{figure}

\begin{figure}
\epsfysize=3in
\epsfbox{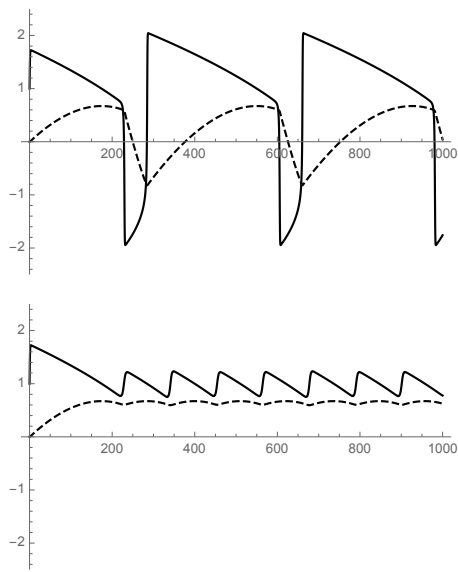}
\caption{Canard implosion in the Van der Pol limit cycle as the bias parameter $a \rightarrow a_{+}(\epsilon) < 1$ approaches the fixed-point stability range. Periodic Van der Pol solutions $x(t)$ (solid) and $y(t)$ (dashed) for $\epsilon = 0.01$ and $a =  0.998739$ (top) and $a =  0.998740$ (bottom).}
\label{fig:vdp_canard}
\end{figure}

Lastly, an important feature of the biased Van der Pol equations \eqref{eq:vdp_a} is that they display canard behavior: canard explosion (Fig.~\ref{fig:vdp_minus_canard}) and canard implosion (Fig.~\ref{fig:vdp_canard}). In Fig.~\ref{fig:vdp_minus_canard}, we see that a small change in the bias parameter $a = -\,0.998740 \rightarrow -\,0.998739$ leads to the appearance of a large-amplitude relaxation oscillation from small-amplitude oscillations about the fixed point, while in Fig.~\ref{fig:vdp_canard}, we see the small change in the bias parameter $a = 0.998739 \rightarrow 0.998740$ leading to the disappearance of large amplitude oscillations in $x(t)$ and $y(t)$ for the case $\epsilon = 0.01$. 

The canard perturbation equation \eqref{eq:canard_eq} for the Van der Pol equations is
\begin{equation}
\epsilon \left[ x \;-\frac{}{} a(\epsilon)\right] \;=\; \pd{\Phi(x,\epsilon)}{x} \left[ \Phi_{0}(x) \;-\frac{}{} \Phi(x,\epsilon)\right],
\end{equation}
where $\Phi_{0}(x) = x - x^{3}/3$ the partial derivatives evaluated at $\epsilon = 0$ are
\begin{equation}
\left. \begin{array}{rcl}
(F_{y0},\; F_{a0}) &=& (-1,\; 0) \\
 && \\
(G_{y0},\; G_{a0}) &=& (0,\; -1)
\end{array} \right\}.
\end{equation}
Here, the lowest-order solution $\Phi_{0}(x)$ has critical points at $x_{c} = \pm 1$, where $\Phi_{0}^{\prime}(x) = 1 - x^{2}$ vanishes. Hence, the lowest-order fixed point $x_{0} = a_{0}$ merges with the critical point $x_{c}$ when $a_{0} = \pm 1$. Because $F_{a0} = 0$, the function $h(x) = 0$ in Eq.~\eqref{eq:canard_eq_1}, while $\Psi_{0}(x) = x + a_{0}$ and $H_{1}(x) = -1$, so that $K_{1}(x) = 1/(x + a_{0}) = \Phi_{1}(x)$.

Next, in Eqs.~\eqref{eq:canard_R2}-\eqref{eq:canard_S1}, we have $R_{2} = -\,K_{1}\,K_{1}^{\prime} = 1/(x+a_{0})^{3}$ and $S_{1} = -1$, so that at $x = a_{0} = \pm 1$, we find the first-order correction $a_{1} = -1/(8\,a_{0}^{3})$, i.e., $a_{1} = -1/8$ for the canard implosion at $a_{0} = 1$, and $a_{1} = 1/8$ for the canard explosion at $a_{0} = -1$. 

For the canard explosion, the calculated critical parameter (truncated at first order) $a_{c}(\epsilon) = -1 + \epsilon/8$ yields $a_{c}(0.01) = -\,0.99875$, which is in excellent agreement with the numerical value $ -\,0.998740...$ shown in Fig.~\ref{fig:vdp_minus_canard}. Because of the symmetry of the Van der Pol model, the calculated critical parameter (truncated at first order) $a_{c}(\epsilon) = 1 - \epsilon/8$ for the canard implosion yields $a_{c}(0.01) = 0.99875$, which is again in excellent agreement with the numerical value $0.998740...$ shown in Fig.~\ref{fig:vdp_canard}. Higher-order corrections to the Van der Pol canard parameter $a_{c}(\epsilon) = 1 - \epsilon/8 - 3\,\epsilon^{2}/32 - 173\,\epsilon^{3}/1024 - \cdots$ can be computed up to arbitrary order \cite{Algaba_2020} but they are not needed in what follows.

\section{\label{sec:CIMA}Chlorine Dioxide Iodine Malonic-Acid (CIMA) Reaction}

Our first example of oscillating chemical reactions is provided by the Chlorine dioxide Iodine Malonic-Acid (CIMA) reactions. Lengyel {\it et al.} \cite{Lengyel_1990,Lengyel_1991,Epstein_1995} proposed the following reduced chemical reactions involving chlorine dioxide, iodine, and malonic acid (MA):
\begin{eqnarray}
{\rm MA} + {\rm I}_{2} & \rightarrow & {\rm IMA} + {\rm I}^{-} + {\rm H}^{+}, \label{eq:I_2} \\
{\rm ClO}_{2} + {\rm I}^{-} & \rightarrow & {\rm ClO}_{2}^{-} + \frac{1}{2}\;{\rm I}_{2}, \label{eq:ClO_2} \\
{\rm ClO}_{2}^{-} + 4\,{\rm I}^{-} + 4\,{\rm H}^{+} & \rightarrow & {\rm Cl}^{-} + 2\,{\rm I}_{2} + 2\,{\rm H}_{2}{\rm O}, \label{eq:ClO_2-}
\end{eqnarray}
By assuming that the concentrations $[{\rm I}_{2}]$, [MA], and $[{\rm ClO}_{2}]$ are constant in time \cite{Lengyel_1991}, the coupled equations for the concentrations $X = [{\rm I}^{-}]$ and $Y = [{\rm ClO}_{2}^{-}]$ satisfy the coupled chemical rate equations
\begin{eqnarray}
\frac{dX}{dt} & = & r_{1} \;-\; k_{2}\,X \;-\; \frac{4\,k_{3}\,X\,Y}{u + X^{2}}, \label{eq:I_dot} \\
\frac{dY}{dt} & = & k_{2}\,X \;-\; \frac{k_{3}\,X\,Y}{u + X^{2}}, \label{eq:ClO2_dot}
\end{eqnarray}
where $(r_{1},k_{2},k_{3},u)$ are positive constants.

We now introduce the normalizations $x = \alpha\,X$, $y = \beta\,Y$, and the dimensionless time is normalized to $\omega^{-1}$. First, we multiply Eq.~\eqref{eq:I_dot} with $\alpha/\omega$ to obtain
\[ \dot{x} \;=\; \frac{\alpha\,r_{1}}{\omega} \;-\; \frac{k_{2}}{\omega}\,x \;-\; \frac{k_{3}\alpha^{2}}{\omega\beta}\;\frac{4\,x\,y}{u\,\alpha^{2} + x^{2}}. \] 
Next, by setting $\alpha = 1/\sqrt{u}$, $\omega = k_{2}$, $\beta = k_{3}/(u\,k_{2})$, and $a = r_{1}/(k_{2}\,\sqrt{u})$, we obtain the dimensionless equation
\begin{equation} 
\dot{x} \;=\; a \;-\; x \;-\; \frac{4\,x\,y}{1 + x^{2}}.
\label{eq:x_dot_CIMA} 
\end{equation}
We multiply Eq.~\eqref{eq:ClO2_dot} with $\beta/\omega$ to obtain
\[ \dot{y} \;=\; \frac{\beta}{\alpha}\;x \;-\; \frac{k_{3}\alpha}{k_{2}}\;\frac{x\,y}{1 + x^{2}}, \]
where we used $\omega = k_{2}$. We now note that, by setting $\alpha\,k_{3}/k_{2} = \beta/\alpha \equiv b = k_{3}/(k_{2}\,\sqrt{u})$, we obtain the dimensionless equation
\begin{equation} 
\dot{y}  \;=\; b\,x \left( 1 \;-\; \frac{y}{1 + x^{2}}\right).
\label{eq:y_dot_CIMA}
\end{equation}
We now show that Eqs.~\eqref{eq:x_dot_CIMA}-\eqref{eq:y_dot_CIMA} can have oscillatory solutions, where $a$ and $b$ are positive dimensionless constants. Under typical laboratory conditions \cite{Awal_Epstein_2020}, we find $0 < a < 35$ and $0 < b < 64$.

\subsection{CIMA Nullclines and Linear Stability}

\begin{figure}
\epsfysize=1.8in
\epsfbox{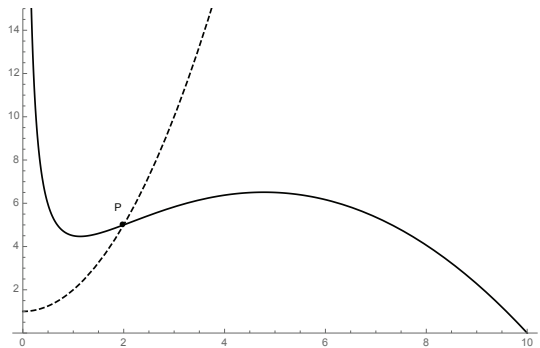}
\caption{Nullclines of the CIMA equations \eqref{eq:x_dot_CIMA}-\eqref{eq:y_dot_CIMA} in the range $0 < x \leq a = 10$. (solid) $x$-nullcline: $y(x) = (a-x)(1 + x^{2})/(4x)$ and (dashed) $y$-nullcline: $y(x) = 1 + x^{2}$. The nullclines intersect at a single fixed point $(x_{0}, y_{0}) = (a/5, 1 + a^{2}/25)$ and the $x$-nullcline has a positive local minimum and a positive local maximum for $a > \sqrt{27}$.}
\label{fig:null_CIMA}
\end{figure}

The nullclines of the CIMA equations \eqref{eq:x_dot_CIMA}-\eqref{eq:y_dot_CIMA} are: \begin{equation}
\left. \begin{array}{rl}
x-{\rm nullcline}: & f(x) = (a-x)(1 + x^{2})/(4x) \\
 & \\
 y-{\rm nullcline}: & g(x) = 1 + x^{2}
 \end{array} \right\},
 \end{equation}
which intersect at a single fixed point $(x_{0}, y_{0}) = (a/5, 1 + a^{2}/25)$. Figure \ref{fig:null_CIMA} shows that the $x$-nullcline has a positive local minimum and a positive local maximum, which exist for $a > \sqrt{27}$. 

\begin{figure}
\epsfysize=1.8in
\epsfbox{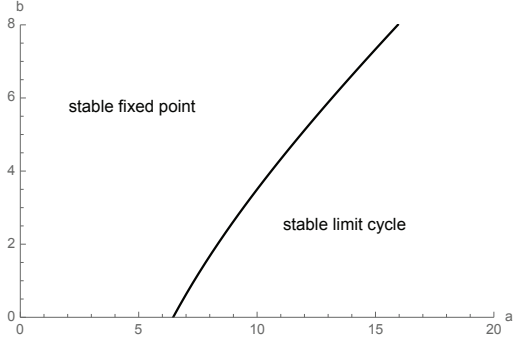}
\caption{Parameter space $(a,b)$ with the solid line corresponding to marginal stability ($\tau = 0$): $b_{c}(a) = 3 a/5 - 25/a$, which is positive for $a > \sqrt{125/3}$. The fixed point $(x_{0}, y_{0} = 1 + x_{0}^{2})$ is stable $(b > b_{c})$ above the solid line, while it  is unstable $(b < b_{c})$ below the solid line, where a stable limit cycle and periodic oscillatory solutions exist.}
\label{fig:b_c}
\end{figure}

The Jacobian matrix at the fixed point $(x_{0}, y_{0})$:
\begin{equation}
{\sf J}(a,b) \;=\; \frac{1}{1 + x_{0}^{2}} \left( \begin{array}{lcr}
3\,x_{0}^{2} - 5 & & -\,4\,x_{0} \\
 & & \\
 2\,b\,x_{0}^{2} & & -\,b\,x_{0}
\end{array} \right)
\end{equation}
has a determinant $\Delta$ and a trace $\tau$ given as
\begin{equation}
\left. \begin{array}{rcl}
\Delta(a,b) &=& 5\,b\,x_{0}/(1 + x_{0}^{2}) \;>\; 0 \\
 && \\
 \tau(a,b) &=& (3\,x_{0}^{2} - b\,x_{0} - 5)/(1 + x_{0}^{2})
 \end{array} \right\}.
\end{equation}
Hence, the fixed point $(x_{0}, y_{0} = 1 + x_{0}^{2})$ is unstable (i.e., the fixed point is repelling) if $\tau > 0$:
\begin{equation}
b \;<\; b_{c} \;=\; 3\,x_{0} \;-\; 5/x_{0} \;=\; 3\,a/5 \;-\; 25/a.
\end{equation}
Figure \ref{fig:b_c} shows the stability parameter $(a,b)$ space, where $b_{c} > 0$ for $a > \sqrt{125/3}$, which also implies that the fixed point is located in the unstable subset of the $x$-nullcline: $x_{D}(a) < x_{0} = a/5 < x_{B}(a)$, when the fixed point is located between the minimum $x_{D}(a)$ and the maximum $x_{B}(a)$ of the $x$-nullcline.

The minimum $x_{D}(a)$ and maximum $x_{B}(a)$ of the $x$-nullcline are the two positive roots of the cubic equation 
\begin{equation}
f^{\prime}(x) \;=\; -\,\left[2\,x^{3} \;+\; a\,\left(1 - x^{2}\right)\right]/(4 x^{2}) \;=\; 0,
\label{eq:cubic_CIMA}
\end{equation}
which are obtained from the general procedure presented in App.~\ref{sec:cubic}. Here, we find
\begin{eqnarray}
x_{D}(a) & = & \frac{a}{3}\left[ \frac{1}{2} \;-\; \cos\left(\frac{\pi}{3} + \frac{\phi(a)}{3}\right) \right], \\
x_{B}(a) & = & \frac{a}{3} \left[ \frac{1}{2} \;+\; \cos\left(\frac{\phi(a)}{3}\right) \right],
\end{eqnarray}
where 
\begin{equation}
\phi(a) \;\equiv\; \arccos(1 - 54/a^{2}). 
\end{equation}
The third root is negative and, therefore, is not relevant (see App.~\ref{sec:cubic}), and we note that the two positive roots $x_{B}(a)$ and $x_{D}(a)$ merge at $a = \sqrt{27}$ (i.e., $\phi = \pi$). 

\subsection{Periodic Oscillatory CIMA Solutions}

For a fixed value of $a > \sqrt{125/3}$, the period of the oscillatory CIMA solution is shortest for $b \simeq b_{c}$ for which the trace $\tau \simeq 0$. In this case, the eigenvalues $\lambda \simeq \pm\,i\;\omega_{c}$ of the CIMA solutions yield a period $T_{c} = 2\pi/\omega_{c}$, where
\begin{equation}
\omega_{c} \;=\;  \sqrt{\Delta_{c}} \;=\; \sqrt{\frac{5\,b_{c}\,x_{0}}{1 + x_{0}^{2}}} \;=\; \sqrt{\frac{15\,a^{2} - 625}{25 + a^{2}}},
\end{equation}
which vanishes at $a = \sqrt{125/3}$. For $a = 10$, we find $\omega_{c} = \sqrt{7}$, while  $\omega_{c} \rightarrow \sqrt{15}$ as $a \rightarrow \infty$. These small-amplitude oscillatory CIMA solutions are 
\begin{equation}
\left. \begin{array}{rcl}
|x(t) - 2| & \leq & \sqrt{\alpha\,(b_{c} - b)} \\
 && \\
 |y(t) - 5| & \leq & \sqrt{\beta\,(b_{c} - b)}
 \end{array} \right\},
 \label{eq:CIMA_supercritical}
 \end{equation}
with numerical constants $(\alpha, \beta) \simeq (2,7)$. Hence, the amplitudes of the oscillatory CIMA reactions vanish as $b \rightarrow b_{c }$, and the scaling \eqref{eq:CIMA_supercritical} is a generic feature of supercritical Hopf bifurcations \cite{Strogatz_2015}.

\begin{figure}
\epsfysize=1.8in
\epsfbox{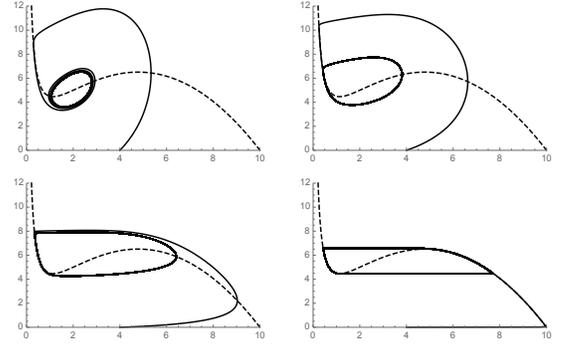}
\caption{Stable limit cycles for $a = 10$ and $b < b_{c} = 3.5$: (top row) $b =$ 3 and 1 (bottom row) $b =$ 0.1 and 0.001. The $x$-nullcline is shown as a dashed curve, and the initial point $(x_{0},y_{0}) = (4,0)$ is used for each orbit. As $b$ approaches zero, the limit cycle approaches an asymptotic limit cycle (see Fig.~\ref{fig:CIMA_vdP}), while for finite values of $b$, the phase-space orbit clearly crosses the $x$-nullcline vertically.}
\label{fig:limit_cycle}
\end{figure}

\begin{figure}
\epsfysize=2in
\epsfbox{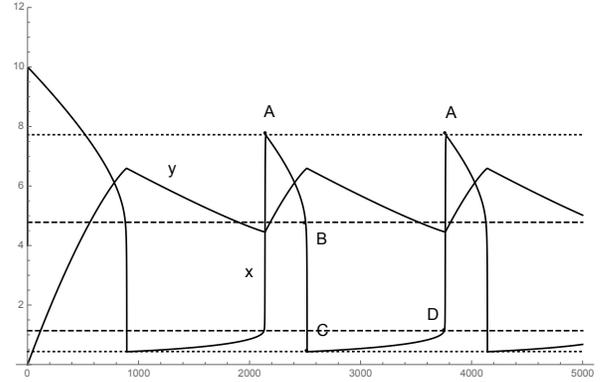}
\caption{Periodic oscillatory CIMA solutions for $x(t)$ and $y(t)$ for $a = 10$ in the asymptotic limit $b = 0.001 \ll 1$. The minimum and maximum (B and D) of the $x$-nullcline are shown as dashed lines, while the minimum and maximum (A and C) of the limit cycle ABCDA are shown as dotted lines.}
\label{fig:period_xy_vdP}
\end{figure}

\begin{figure}
\epsfysize=2in
\epsfbox{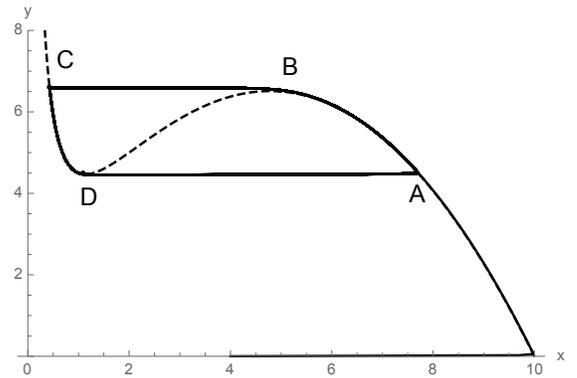}
\caption{Phase portrait $y(t)$ versus $x(t)$ for the periodic CIMA solutions shown in Fig.~\ref{fig:period_xy_vdP}, with a well-defined asymptotic limit cycle $ABCDA$ and the $x$-nullcline is shown as a dashed curve. The time scales for the fast horizontal orbits $B \rightarrow C$ and $D \rightarrow A$ are much shorter that the slow orbits on the $x$-nullcline (shown as a dashed curve) $A \rightarrow B$ and $C \rightarrow D$.}
\label{fig:CIMA_vdP}
\end{figure}

\begin{figure}
\epsfysize=2in
\epsfbox{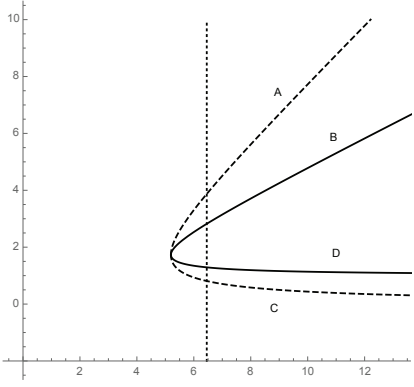}
\caption{Limit-cycle functions $x_{A}(a)$, $x_{B}(a)$, $x_{C}(a)$, and $x_{D}(a)$ in the range $a \geq \sqrt{27}$ (where they merge at $\sqrt{3}$). The vertical dotted line  $(b_{c} = 0)$ at $a = \sqrt{125/3}$ is used to indicate that the asymptotic limit cycle ABCDA is stable $(0 < b \;=\; \epsilon \ll b_{c})$ only for $a > \sqrt{125/3}$.}
\label{fig:x_ABCD}
\end{figure}

For a fixed value of $a > \sqrt{125/3}$, the longest periods are found in the asymptotic limit: $b = \epsilon \ll 1$. The limit-cycle maximum $x_{A}(a)$ and minimum $x_{C}(a)$ (see Figs.~\ref{fig:period_xy_vdP}-\ref{fig:x_ABCD}) are solutions of the equations
\begin{eqnarray*}
y_{D} &=& (a - x_{D})\,(1 + x_{D}^{2})/(4 x_{D}) \\
 &=& (a - x_{A})\,(1 + x_{A}^{2})/(4 x_{A}), \\
y_{B} &=& (a - x_{B})\,(1 + x_{B}^{2})/(4 x_{B}) \\
 &=& (a - x_{C})\,(1 + x_{C}^{2})/(4 x_{C}),
\end{eqnarray*}
which can be rewritten as cubic equations (with $x_{B} \neq 0 \neq x_{D}$)
\begin{eqnarray}
x_{A}^{3} - a\,x_{A}^{2} + \left(\frac{a}{x_{D}} + a\,x_{D} - x_{D}^{2}\right)\,x_{A} - a &=& 0, \label{eq:xA_eq} \\
x_{C}^{3} - a\,x_{C}^{2} + \left(\frac{a}{x_{B}} + a\,x_{B} - x_{B}^{2}\right)\,x_{C} - a &=& 0, \label{eq:xC_eq}
\end{eqnarray}
where $2\,x_{D}^{3}  - a\,(x_{D}^{2} - 1) = 0 = 2\,x_{B}^{3}  - a\,(x_{B}^{2} - 1)$. Since $x_{D}$ and $x_{B}$ are double roots of their respective equations, we divide them by $(x_{A} - x_{D})^{2}$ and $(x_{C} - x_{B})^{2}$, respectively, and we find
\begin{eqnarray}
x_{A}(a) &=& a - 2\,x_{D}(a) = \frac{2a}{3}\left[ 1 + \cos\left(\frac{\pi}{3} + \frac{\phi(a)}{3}\right) \right].
\label{eq:xA_def} \\
x_{C}(a) &=& a - 2\,x_{B}(a) = \frac{2a}{3}\left[ 1 - \cos\left(\frac{\phi(a)}{3}\right) \right].
\label{eq:xC_def}
\end{eqnarray}
Figure \ref{fig:x_ABCD} shows plots of $(x_{A},x_{B},x_{C},x_{D})$ as functions of $a$, where are seen to merge at $a = \sqrt{27}$.

\begin{figure}
\epsfysize=2in
\epsfbox{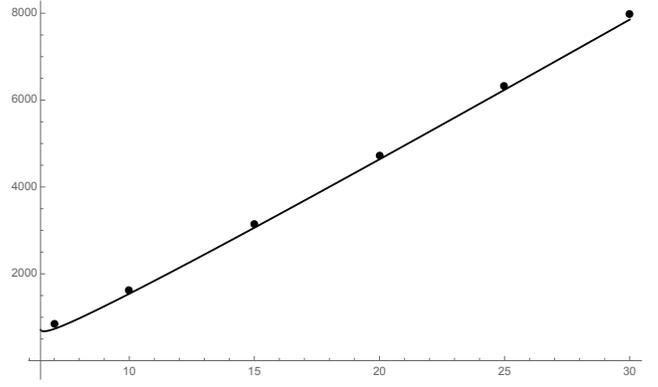}
\caption{Plot of the asymptotic limit-cycle period $\epsilon\,T_{\rm CIMA}(a)$ determined by Eq.~\eqref{eq:P_th} in the range $a \geq \sqrt{125/3}$, with dots representing the numerical CIMA periods for various values of $a$ in the asymptotic limit $b = \epsilon = 0.001 \ll 1$.}
\label{fig:P_th}
\end{figure}

The asymptotic limit \eqref{eq:vdP_period} of the CIMA period is expressed as
\begin{equation}
\epsilon\,T_{\rm CIMA}(a) \;=\;  \int_{x_{A}(a)}^{x_{B}(a)} \frac{f^{\prime}(x)\;dx}{G(x,f(x))} \;+\; \int_{x_{C}(a)}^{x_{D}(a)} \frac{f^{\prime}(x)\;dx}{G(x,f(x))},
\label{eq:period_P}
\end{equation}
where the integrand
\begin{eqnarray} 
\frac{f^{\prime}(x)}{G(x, f(x))} &=& \frac{[a\,(1 - x^{2}) + 2\,x^{3}]}{(a - 5x)\,x^{2}} \nonumber \\
 &=& -\,\frac{2}{5} + \frac{5}{a\,x} + \frac{1}{x^{2}} + \frac{b_{c}(a)}{5\,x - a},
\end{eqnarray}
has been decomposed in terms of partial fractions. The asymptotic limit-cycle CIMA period \eqref{eq:period_P} is, therefore, explicitly expressed as
\begin{eqnarray}
\epsilon\,T_{\rm CIMA}(a) & = & -\,\frac{2}{5} \left( 3\,x_{B} + 3\,x_{D} - 2\,a\right) + \frac{5}{a} \ln\left( \frac{x_{B}\;x_{D}}{x_{A}\,x_{C}}\right)  \nonumber \\
 &&-\; \left[ \left(\frac{1}{x_{B}} - \frac{1}{x_{A}}\right) + \left(\frac{1}{x_{D}} - \frac{1}{x_{C}}\right)\right] \nonumber \\
  &&+\; \frac{1}{5}\,b_{c}(a) \ln\left[ \frac{(5 x_{B} - a)\,(5 x_{D} - a)}{(5 x_{A} - a)\,(5 x_{C} - a)} \right].
  \label{eq:P_th}
\end{eqnarray}
Figure \ref{fig:P_th} shows a plot of Eq.~\eqref{eq:P_th} in the range $a \geq \sqrt{125/3}$, with dots representing the numerical CIMA periods for various values of $a$. For $a = 10$ and $\epsilon = 0.001$, for example, the asymptotic period $T_{\rm CIMA}(a,\epsilon) = 1541$ is only 5\% below the numerical value $T_{num}(a,\epsilon) = 1625$. The error between the asymptotic period $T_{\rm CIMA}(a,\epsilon)$ and the numerical value $T_{num}(a,\epsilon)$ falls below 2\% as $a > 15$.

In the asymptotic limit $a \rightarrow \infty$, we find $\phi(a) \rightarrow 6\sqrt{3}/a$, using the Taylor expansion for $\arccos(1 - x^{2}) \simeq \sqrt{2}\,x$, so that $x_{A} \rightarrow a$, $x_{B} \rightarrow a/2$, $x_{C} \rightarrow 4/a$, and $x_{D} \rightarrow 1$, which yields a linear dependence in $a$, with a slope $\epsilon\,T_{\rm CIMA}(a,\epsilon)/a \rightarrow (3/25)\left[ (15/4) - \ln(8/3) \right] \simeq 0.332$.

\subsection{Canard explosion for the CIMA model}

Since the fixed point $x_{0} = a/5$ reaches the minimum $x_{D}(a)$ of the $x$-nullcline when $a = \sqrt{125/3} = 6.45497...$, we can expect a canard explosion in the vicinity of $a_{c} \simeq \sqrt{125/3}$, when $b = \epsilon \ll 1$. The fixed point, however, can never reach the maximum of the $x$-nullcline so a canard implosion is not possible for the CIMA model. 

\begin{figure}
\epsfysize=2in
\epsfbox{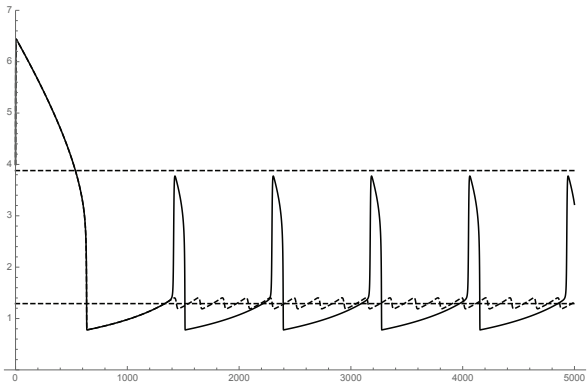}
\caption{Canard explosion  in the CIMA model (for $b = 0.001 \ll 1$): $a =  6.460$ (dashed) $\rightarrow$ $a = 6.461$ (solid), where a small-amplitude oscillation (dashed) about the minimum of the $x$-nullcline explodes into a large-amplitude oscillation (solid).}
\label{fig:CIMA_canard}
\end{figure}

Here, the invariant manifold $y = \Phi(x,\epsilon)$ for the CIMA equations yields the canard perturbation equation
\begin{equation}
\epsilon \left( x - \frac{x\,\Phi(x,\epsilon)}{1+x^{2}}\right) = \pd{\Phi(x,\epsilon)}{x} \left( a - x - \frac{4x\,\Phi(x,\epsilon)}{1+x^{2}}\right),
\end{equation}
where, defining the function $\psi(x) = x/(1 + x^{2})$, we have the partial derivatives evaluated at $\epsilon = 0$:
\begin{equation}
\left. \begin{array}{rcl}
F_{y0} &=& -4\,\psi(x) \\
F_{a0} &=& 1 \\
G_{y0} &=& -\,\psi(x) \\ 
G_{a0} &=& 0
\end{array} \right\}.
\end{equation}
Here, the lowest-order solution $\Phi_{0}(x) = \frac{1}{4}\,(a_{0} - x)/\psi(x)$ has a minimum at $x_{D} = a_{0}/5 = \sqrt{5/3}$, which coincides with the fixed point at $a_{0} = \sqrt{125/3}$. At first order, the first-order solution \eqref{eq:canard_1_sol}: $\Phi_{1}(x) = K_{1}(x) - h(x)\,a_{1}$ is expressed in terms of the functions $K_{1}(x)$ and $h(x)$ defined in Eq.~\eqref{eq:canard_eq_1}:
\[ K_{1}(x) \;=\; \frac{5\,x\,(1 + x^{2})}{4\,(2 x^{2} - \sqrt{15}\,x - 5)} \]
and $h(x) = -1/[4\,\psi(x)] = -\frac{1}{4}\,(1 + x^{2})/x$.

At second order, we can now evaluate the functions $R_{2}(x)$ and $S_{1}(x)$, defined in Eqs.~\eqref{eq:canard_R2}-\eqref{eq:canard_S1}, as
\begin{equation}
\left. \begin{array}{rcl}
R_{2}(x) & = & \psi(x)\,K_{1}(x) \left[ 1 \;-\; 4\,K_{1}^{\prime}(x) \right] \\
 && \\
S_{1}(x) & = & -\,1/4 \;-\; K_{1}(x)\,\psi^{\prime}(x)/\psi(x)
\end{array} \right\}.
\end{equation}
When these functions are evaluated at $x_{D} = a_{0}/5 = \sqrt{5/3}$, we find $R_{2}(x_{D}) = -15/8$ and $S_{1}(x_{D}) = -3/8$, so that $a_{1} = R_{2}(x_{D})/S_{1}(x_{D}) = 5$.

By substituting the value $\epsilon = 0.001$ in the first-order truncated expression for the critical canard parameter for the CIMA model
\begin{equation}
a_{c}(\epsilon) \;=\; a_{0} \;+\; a_{1}\,\epsilon \;=\; \sqrt{125/3} \;+\; 5\,\epsilon 
\label{eq:a_CIMA}
\end{equation}
for the canard explosion near the minimum of the $x$-nullcline of the CIMA, we obtain $a_{c}(0.001) = 6.45997$, which is in excellent agreement with the numerical value $6.460$ shown in Fig.~\ref{fig:CIMA_canard}. If needed, higher-order corrections to Eq.~\eqref{eq:a_CIMA} can be calculated from Eq.~\eqref{eq:canard_01}. A similar canard-explosion analysis of the CIMA equations, based on the Krupa-Szmolyan \cite{Krupa_2001} perturbation analysis, was recently performed by Awal and Epstein \cite{Awal_Epstein_2020}, which yielded results that are identical to our perturbation analysis.

\section{\label{sec:Oregonator}Oregonator Models of the BZ Reaction}

Our second example of oscillatory chemical reactions is provided by the Belousov-Zhabotinsky (BZ) reactions. The Oregonator model \cite{FKN_1972,Field_Noyes_1974} of the BZ reactions is expressed in terms of the three coupled chemical rate equations
\begin{eqnarray}
\dot{X} & = & k_{1}A\,Y \;-\; k_{2}\,X\,Y \;+\; k_{3}A\,X \;-\; 2 k_{4}\,X^{2}, \label{eq:X_dot} \\
\dot{Y} & = & -\,k_{1}A\,Y \;-\; k_{2}\,X\,Y \;+\; \frac{1}{2}\,k_{c}B\;\sigma\,Z, \label{eq:Y_dot} \\
\dot{Z} & = & 2\,k_{3}A\,X \;-\; k_{c}B\,Z, \label{eq:Z_dot}
\end{eqnarray}
where the important chemical species are $X = [{\rm HBrO}_{2}]$, $Y = [{\rm Br}^{-}]$, $Z = 2\,[{\rm Ce}^{4+}]$, $A = [{\rm BrO}_{3}^{-}]$, $B = [{\rm CH}_{2}({\rm COOH})_{2}]$, the rates $(k_{1},k_{2},k_{3},k_{4},k_{c})$ are all positive, and the stoichiometric ratio $\sigma > 0$ is a free parameter \cite{FKN_1972}. We note here that $A$ and $B$ are in excess in these reactions and do not evolve over the time scales of interest.

\subsection{Oregonator-3 equations}

We obtain the following dimensionless Oregonator equations by introducing the normalizations $x = \alpha\,X$, $y = \beta\,Y$, and $z = \gamma\,Z$, with dimensionless time normalized to $\omega^{-1}$, to obtain the dimensionless Oregonator-3 equations
\begin{eqnarray}
\dot{x} & = & y \;(q - x) \;+\; x\,(1 - x), \label{eq:x_prime} \\
\epsilon\,\dot{y} & = & \sigma\,z \;-\; y\,(q + x), \label{eq:y_prime} \\
\dot{z} & = & \delta\,(x - z), \label{eq:z_prime}
\end{eqnarray}
where the dimensionless constants $(q, \epsilon, \delta)$, which are small and positive, and the normalization factors $(\alpha,\beta,\gamma,\omega)$ are expressed in terms of the Oregonator parameters $(k_{1},k_{2},k_{3},k_{4},k_{c})$ and $(A,B)$:\begin{equation}
\left. \begin{array}{rcl}
\alpha & = & 2\,k_{4}/(k_{3}A) \\
\beta & = & k_{2}/(k_{3}A) \\
\gamma & = & k_{4}k_{c}B/(k_{3}A)^{2} \\
\omega & = & k_{3}A
\end{array} \right\},
\label{eq:OBZ}
\end{equation}
with the model parameters
\begin{equation}
\left. \begin{array}{rcl}
q & = & 2\,k_{4}\,k_{1}/(k_{3}\,k_{2}) \\
\epsilon & = & 2\,k_{4}/k_{2}  \\
\delta & = & k_{c}B/(k_{3}A) \\
\end{array} \right\}.
\end{equation}
Other normalizations have appeared in the literature \cite{Tyson_Fife_1980,Brons_1991,Field_2007}. The standard normalization of Field's Scholarpedia review \cite{Field_2007} uses the same normalization \eqref{eq:OBZ} for $(x,y,z)$ but uses a different dimensionless time based on $\ov{\omega} = \delta\,\omega$, so that $(\ov{\epsilon},\ov{\epsilon}^{\prime}) = (\delta,\delta\,\epsilon)$. Adapted from Field's Oregonator model \cite{Field_2007}, we use the following parameter values: 
\begin{equation}
(q, \epsilon, \delta) \;=\; \left(7.62 \times 10^{-5},\; 2 \times 10^{-3},\; 1 \times 10^{-3}\right),
\label{eq:O_par}
\end{equation}
where the value $\delta = 0.001$ is used instead of $0.0099$ in order to improve our asymptotic analysis of the Oregonator-2 model.

\begin{figure}
\epsfysize=2in
\epsfbox{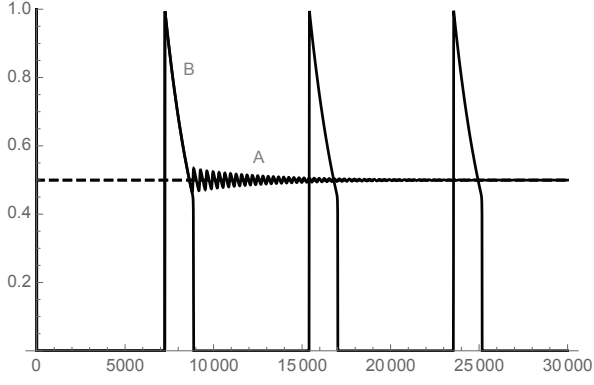}
\caption{Canard explosion in the numerical Oregonator-3 solution $x_{3}(\tau)$ for $\sigma = 0.50047$ (A) and $\sigma = 0.50048$ (B), from a stable fixed-point solution (A) to a large-amplitude relaxation oscillation (B).}
\label{fig:xyz_1}
\end{figure}

The Oregonator-3 equations \eqref{eq:x_prime}-\eqref{eq:z_prime} have a fixed point $(x_{0},y_{0},z_{0})$, where $z_{0} = x_{0} = x_{+}$ and $y_{0} = \sigma x_{+}/(q + x_{+}) = x_{+}(1 - x_{+})/(x_{+} - q)$, where the positive root of the quadratic equation 
\begin{equation}
x^{2} \;+\; (\sigma + q - 1)\,x \;-\; (1 + \sigma)\,q \;=\; 0
\label{eq:O3_fp}
\end{equation}
is expressed as
\begin{equation} 
x_{+}(\sigma,q) = \frac{1}{2}\left[ A(\sigma,q) \;+\frac{}{} B(\sigma,q)\right], 
\label{eq:xs_O3}
\end{equation} 
with
\begin{equation}
\left. \begin{array}{rcl}
A(\sigma,q) &=& 1 - \sigma - q \\
 && \\
 B(\sigma,q) &=& \sqrt{(\sigma + q - 1)^{2} + 4\,(1 + \sigma)\,q} 
 \end{array} \right\}.
 \label{eq:AB_fixed}
 \end{equation}
 The negative root $x_{-}(\sigma,q) = \frac{1}{2}[ A(\sigma,q) + B(\sigma,q)]$ will be used below. For the model parameters used here, a limit cycle appears in $(x,y,z)$-space when $\sigma > 0.50047...$. In Fig.~\ref{fig:xyz_1}, the numerical solutions for $x_{3}(\tau)$ for the Oregonator-3 model \eqref{eq:x_prime}-\eqref{eq:z_prime} are shown for the parameter values \eqref{eq:O_par} and $\sigma = 0.50047$ (A) and $\sigma = 0.50048$ (B), which shows a canard explosion to a large-amplitude stable limit cycle. 
 
\begin{figure}
\epsfysize=1.8in 
\epsfbox{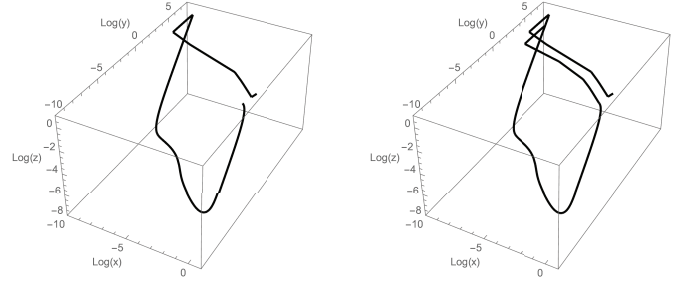}
\caption{Canard-explosion behavior in the numerical Oregonator-3 solutions shown in 3D logarithmic space $(\ln x, \ln y, \ln z)$ for $\sigma = 0.50047$ (left) and $\sigma = 0.50048$ (right). We note that for $\sigma < 0.50048...$, the numerical solutions settle to their steady-state values, while for $\sigma \geq 0.50048..$, a stable limit cycle suddenly appears.}
\label{fig:xyz_3D}
\end{figure}

 Figure \ref{fig:xyz_3D} shows these numerical solutions in 3D logarithmic space $(\ln x, \ln y, \ln z)$. Here, the canard explosion involves large-amplitude relaxation oscillations in the $y$-variable (i.e., the $[{\rm Br}^{-}]$ ion species), as can be seen in Fig.~\ref{fig:xyz_32}. The analysis of canard explosions can also be carried out in three dimensions (e.g., see Ref.~\cite{Szmolyan_2001}) but this analysis is outside the scope of our work.

\subsection{Oregonator-2 equations}

When $y$ is slowly varying, with $\epsilon \ll 1$ in Eq.~\eqref{eq:y_prime}, then $\epsilon\,\dot{y}$ may be taken to be zero in Eq.~\eqref{eq:y_prime} and, thus, we may use the constraint equation 
\begin{equation}
y \;=\; \sigma\,z/(q + x). 
\label{eq:y_constraint}
\end{equation}
This approximation must be checked afterward by comparing numerical solutions of the three-dimensional Oregonator-3 model and the two-dimensional Oregonator-2 model. We note that the Oregonator-2 model derived from the dimensional Oregonator equations \eqref{eq:X_dot}-\eqref{eq:Z_dot} depends on the normalization used to derive the dimensionless Oregonator-3 equations \cite{Brons_footnote}.

By substituting the constraint \eqref{eq:y_constraint} into Eq.~\eqref{eq:x_prime}, we obtain the two-field Oregonator-2 equations
\begin{eqnarray}
\dot{x} & = & x\,(1 - x) \;+\; \left(\frac{q - x}{q + x}\right) \sigma\,z \;=\; F(x,z), \label{eq:x2_prime} \\
\dot{z} & = & \delta\,(x - z) \;=\; \delta\,G(x,z), \label{eq:z2_prime} 
\end{eqnarray}
which yield the $x$-nullcline and $z$-nullcline, respectively:
\begin{eqnarray}
z(x) & = & \frac{x\,(1 - x)\,(x + q)}{\sigma\;(x - q)} \;=\; \sigma^{-1}\,\varphi(x,q), \label{eq:x_null} \\
z(x) & = & x. \label{eq:z_null}
\end{eqnarray}
These nullclines intersect at $x = 0$ and $x = x_{+}(\sigma,q)$ defined by Eq.~\eqref{eq:xs_O3}.

We note that the function $\varphi(x,q)$ is positive in the range $q < x < 1$ and it has a local minimum $x_{D}(q)$ and a local maximum $x_{B}(q)$ for $0 \leq q \leq Q$. The minimum and maximum are positive roots of the cubic equation
\begin{equation} 
2\,x^{3} \;-\; (1 + 2\,q)\;x^{2} \;+\; 2\,q\,(1 - q)\;x \;+\; q^{2} \;=\; 0.
\label{eq:cubic_O2}
\end{equation}
Using the method described in App.~\ref{sec:cubic}, these positive roots $0 < x_{D}(q) \leq x_{B}(q)$ are
\begin{eqnarray}
x_{D}(q) & = & \frac{1}{6}\,(1 + 2\,q) - \frac{1}{3}\,(1 - 4\,q)\cos\left(\frac{\pi}{3} + \frac{\phi(q)}{3}\right), \label{eq:x1_min} \\
x_{B}(q) & = & \frac{1}{6}\,(1 + 2\,q) + \frac{1}{3}\,(1 - 4\,q)\cos\left(\frac{\phi(q)}{3}\right), \label{eq:x2_max}
\end{eqnarray}
where 
\[ \phi(q) \;=\; \arccos[(1 - 12\,q - 60\,q^{2} + 44\,q^{3})/(1 - 4\,q)^{3}]. \]
The roots \eqref{eq:x1_min}-\eqref{eq:x2_max} merge when $\phi(Q) = \arccos(-1) = \pi$, where $Q = -\,(1/5) + (6/5)\;\sinh\left[(1/3)\;{\rm arcsinh}(3/4) \right] = 0.0797...$, which is the only real root of the cubic equation $10\,Q^{3} + 6\,Q^{2} + 12\,Q - 1$. 

\begin{figure}
\epsfysize=1.8in
\epsfbox{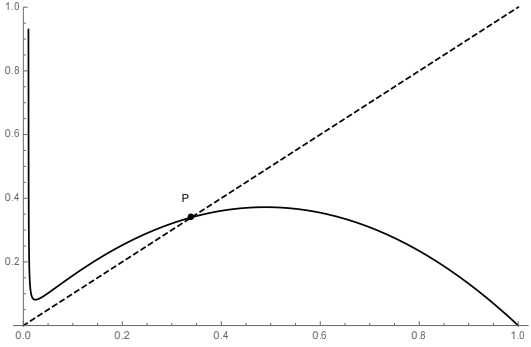}
\caption{Plots of $x$-nullcline $z = \sigma^{-1}\varphi(x,q)$ (solid) and the $z$-nullcline $z = x$ (dashed) in the range $0 < x < 1$ for $q = 0.01 < Q$ and $\sigma = 0.7$. The fixed point $(x_{s},z_{s} = x_{s})$ at point P is at the intersection of the two nullclines.}
\label{fig:Oregon_fixed}
\end{figure}

The linear stability of the fixed point $(x_{0} = x_{+} = z_{0})$ is investigated in terms of the Jacobian matrix equation \eqref{eq:Jac}, where the trace $\tau(\sigma;q,\delta) = F_{x0} - \delta$ and determinant $\Delta = -\,\delta\;(F_{x0}+F_{z0})$ are defined in terms of
 \be 
F_{x0}(\sigma,q) & = & 1 - 2\,x_{+}(\sigma,q) - \frac{2\sigma q\,x_{+}(\sigma,q)}{[q + x_{+}(\sigma,q)]^{2}}, \\
 F_{z0}(\sigma,q) & = & \sigma \left(\frac{q - x_{+}(\sigma,q)}{q + x_{+}(\sigma,q)}\right).
 \ee
For the model parameters \eqref{eq:O_par} used here, a limit cycle is stable in $(x,z)$-space when the trace $\tau$ is positive between $\sigma_{s} = 0.500729$ and $\sigma_{u} = 2.41175$. 

Figure \ref{fig:xz_1} shows that a canard explosion occurs as $\sigma$ crosses the threshold value $\sigma > 0.50047...$ We note that that fixed point reaches the maximum of the $x$-nullcline $\varphi(x,q)/\sigma$ when $\sigma_{B} = \varphi(x_{B},q)/x_{B} = 0.500229$ and the minimum of $x$-nullcline $\varphi(x,q)/\sigma$ when $\sigma_{D} = \varphi(x_{D},q)/x_{D} = 2.41346$. We note that the fixed-point P moves to the left as $\sigma$ increases. Hence, a canard explosion is expected to occur near the maximum $x_{B}(q)$ while a canard implosion is expected to occur near the minimum $x_{D}(q)$.

\begin{figure}
\epsfysize=1.8in
\epsfbox{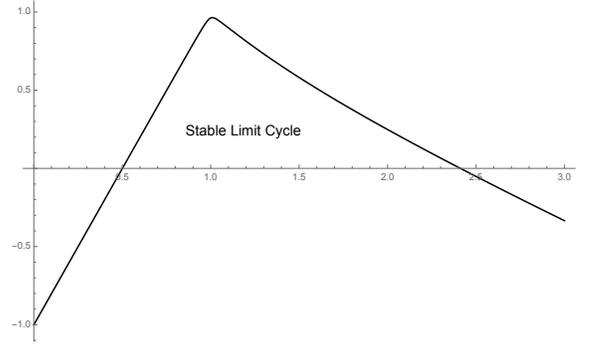}
\caption{Plot of the trace $\tau(\sigma;q,\delta)$ as a function of $\sigma$ for $q = 7.62 \times 10^{-5}$ and $\delta = 0.001$. A limit cycle is stable in the range $\sigma_{s} < \sigma < \sigma_{u}$, where $\sigma_{s} = 0.500729$ and $\sigma_{u} = 2.41175$. We note that $\sigma_{B} < \sigma_{s}$, i.e., the limit cycle becomes stable after the fixed point has reached the minimum of the $x$-nullcline, and $\sigma_{u} < \sigma_{D}$, i.e., the limit cycle becomes unstable before the fixed point has reached the maximum of the $x$-nullcline.}
\label{fig:Oregon_stable}
\end{figure}

\subsection{Asymptotic Period for the Oregonator-2 Model}

\begin{figure}
\epsfysize=1.8in
\epsfbox{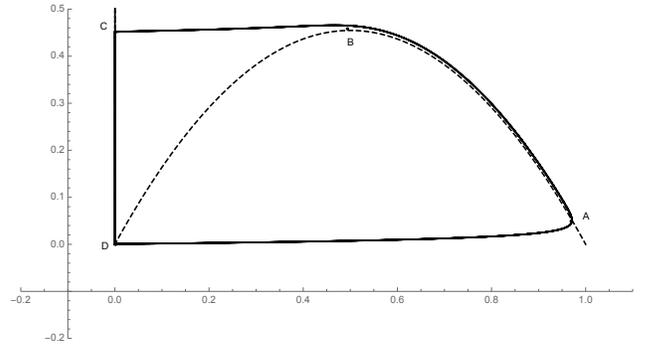}
\caption{Calculation of the period of the asymptotic limit cycle ABCDA for the Oregonator-2 oscillatory solution for $\sigma = 0.55$.}
\label{fig:vdP_period}
\end{figure}

The asymptotic approximation of the Oregonator-2 period is expressed as
\begin{eqnarray}
\delta\,T_{{\rm O}_{2}}(q,\sigma,\delta) &=& \int_{x_{A}(q)}^{x_{B}(q)}\frac{\varphi_{x}(x,q)\;dx}{\sigma\,x - \varphi(x,q)} \nonumber \\
 &&+\; \int_{x_{C}(q)}^{x_{D}(q)}\frac{\varphi_{x}(x,q)\;dx}{\sigma\,x - \varphi(x,q)}, 
\label{eq:T_vdP_O2}
\end{eqnarray}
where we consider the asymptotic limit $\delta \ll 1$ and the integrand is calculated by first using the $x$-nullcline: $z = \sigma^{-1}\varphi(x,q)$ and taking the time derivative $dz/dt = \sigma^{-1}\varphi_{x}\;dx/dt$. Next, we use the $z$-equation $dz/dt = 
\delta\,(x - z)$ and substitute the $x$-nullcline $dz/dt = \delta\,[x - \sigma^{-1}\,\varphi(x,q)]$. By comparing the two $\dot{z}$-equations, we find $\delta\,dt = \varphi_{x}(x,q)\,dx/[\sigma x - \varphi(x,q)]$.

\begin{figure}
\epsfysize=1.8in
\epsfbox{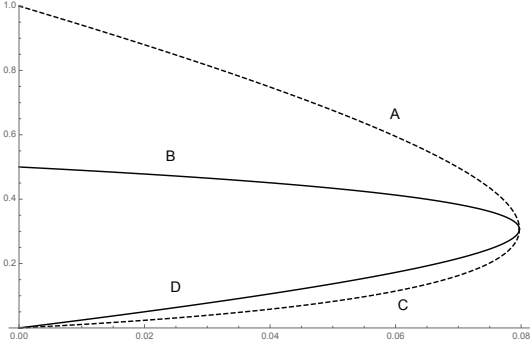}
\caption{Plots of the Oregonator-2 limit cycle ABCDA roots $x_{A}(q)$, $x_{B}(q)$, $x_{C}(q)$, and $x_{D}(q)$}
\label{fig:xq_O2}
\end{figure}

In Eq.~\eqref{eq:T_vdP_O2}, $x_{D}(q)$ and $x_{B}(q)$ are the local minimum and local maximum of the $x$-nullcline, while 
\begin{equation}
\left. \begin{array}{rcl}
x_{A}(q) &=& (1 - q) - 2\,x_{D}(q) \\
 && \\
x_{C}(q) &=& (1 - q) - 2\,x_{B}(q) 
\end{array} \right\}
\end{equation}
 are the single roots of the cubic equations $\sigma\,z_{D} = \varphi(x,q)$ and $\sigma\,z_{B} = \varphi(x,q)$, respectively. These four roots are shown in Fig.~\ref{fig:xq_O2}, where they are seen to merge at $q = Q \equiv -\frac{1}{5} + \frac{6}{5} \sinh[\frac{1}{3}{\rm arcsinh}(\frac{3}{4})] \simeq 0.08$. We also note that the asymptotic period \eqref{eq:T_vdP_O2} goes to zero as $q \rightarrow Q$.

\begin{figure}
\epsfysize=2in
\epsfbox{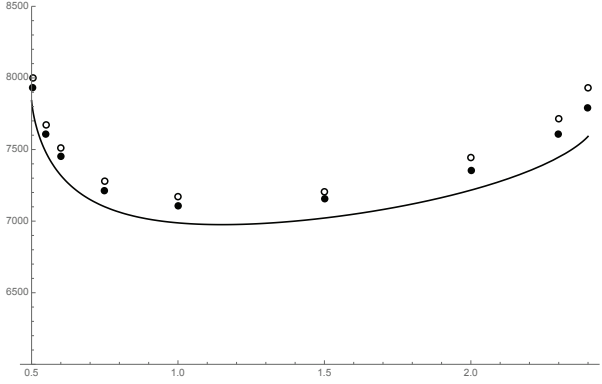}
\caption{Oregonator-2 (dots) and Oregonator-3 (open circles) numerical periods compared to the asymptotic approximation (solid) $T_{{\rm O}_{2}}(q,\delta,\sigma)$, defined by Eq.~\eqref{eq:O_2_period}, in the range $0.505 \leq \sigma \leq 2.4$ for $\delta = 0.001$ and $q = 7.62 \times 10^{-5}$. As expected, the Oregonator-3 periods are slightly longer than the Oregonator-2 periods and the asymptotic Oregonator-2 period captures faithfully the parametric dependence of the period on the stoichiometric ratio $\sigma$.}
\label{fig:Oregonator_2_Period}
\end{figure}

We now evaluate the integrals in Eq.~\eqref{eq:T_vdP_O2} by using the partial-fraction decomposition
\begin{eqnarray}
\frac{\varphi_{x}(x,q)}{\sigma\,x - \varphi(x,q)} &=&  \frac{E_{+}(\sigma,q)}{x - x_{+}(\sigma,q)} + \frac{E_{-}(\sigma,q)}{x - x_{-}(\sigma,q)} \nonumber \\
 &&-\; \frac{1}{(1 + \sigma)\,x} \;+\; \frac{1}{x - q},
\end{eqnarray}
where $x_{\pm}(\sigma,q) = \frac{1}{2}\,(A \pm B)$ are roots of the fixed-point equation \eqref{eq:O3_fp}, with $A(\sigma,q)$ and $B(\sigma,q)$ defined in Eq.~\eqref{eq:AB_fixed}, and 
\[ E_{\pm}(\sigma,q) \;=\; \frac{1}{2}\;\left[-\,C(\sigma) \;\pm\frac{}{} D(\sigma,q)/B(\sigma,q)\right], \]
with $C = (2 + 3 \sigma)/(1 + \sigma)$ and $D = (3 - q + \sigma)\sigma/(1 + \sigma)$. The asymptotic approximation \eqref{eq:T_vdP_O2} of the Oregonator-2 period can therefore be exactly evaluated as
\begin{eqnarray}
\delta\,T_{{\rm O}_{2}}(q,\sigma,\delta) &=& E_{+}(\sigma,q) \;\ln\left[  \frac{(x_{B}-x_{+})\;(x_{D}-x_{+})}{(x_{A}-x_{+})\;(x_{C}-x_{+})}\right] \nonumber \\
  &&+\; E_{-}(\sigma,q) \;\ln\left[  \frac{(x_{B}-x_{-})\;(x_{D}-x_{-})}{(x_{A}-x_{-})\;(x_{C}-x_{-})}\right] \nonumber \\
 &&-\; \frac{1}{(1 + \sigma)} \ln\left( \frac{x_{B}\;x_{D}}{x_{A}\;x_{C}}\right) \nonumber \\
  &&+\; \ln\left[  \frac{(x_{B}-q)\;(x_{D}-q)}{(x_{A}-q)\;(x_{C}-q)}\right],
\label{eq:O_2_period} 
\end{eqnarray}
which is shown in Fig.~\ref{fig:Oregonator_2_Period}. The numerical periods of the Oregonator-2 (dots) and Oregonator-3 (open circles) equations are also shown in Fig.~\ref{fig:Oregonator_2_Period}, which show excellent agreements with asymptotic approximation \eqref{eq:O_2_period} of the Oregonator-2 period. For example, using $\sigma = 1$, the Oregonator-2 numerical period is $7110$, the Oregonator-3 numerical period is $7160$, and the asymptotic period \eqref{eq:O_2_period} is $6988$, which is just 1.7\% lower than the Oregonator-2 period and 2.4\% lower than the Oregonator-3 period. 

Lastly, we note that since $x_{+}(\sigma,q)$ represents the fixed point of the Oregonator-2 equations, the asymptotic period \eqref{eq:O_2_period} becomes infinite when we reach $x_{+} = x_{B}$ at $\sigma = \sigma_{B}$ (i.e., the maximum of the $x$-nullcline) or $x_{+} = x_{D}$ at $\sigma = \sigma_{D}$ (i.e., the minimum of the $x$-nullcline).

\subsection{Canard behavior in the Oregonator-2 model}

\begin{figure}
\epsfysize=1.8in
\epsfbox{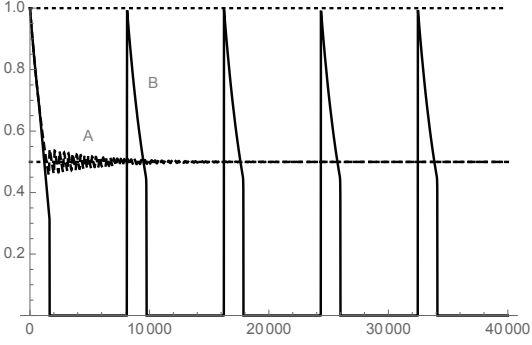}
\caption{Canard explosion in the numerical Oregonator-2 solution $x_{2}(\tau)$ for $\sigma = 0.500478$ (A) and $\sigma = 0.500479$ (B), for the parameters \eqref{eq:O_par}. Note that the amplitude of the $x$-solution is limited by $x = 1$.}
\label{fig:xz_1}
\end{figure}

 \begin{figure}
\epsfysize=1.8in 
\epsfbox{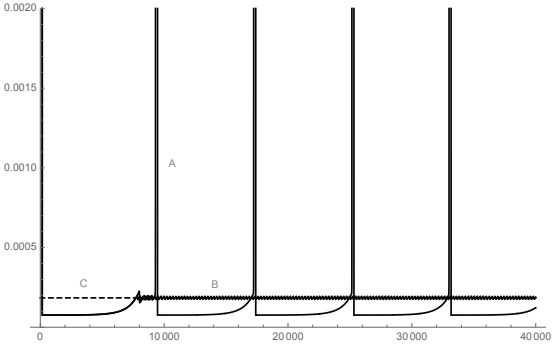}
\caption{Canard implosion in the numerical Oregonator-2 solution $x_{2}(\tau)$ shown for $\sigma = 2.4106$ (A) and $\sigma = 2.4109$ (B). Note that the relaxation-oscillation solution (A) still reaches a maximum value at $x = 1$ and that the small-amplitude periodic solution (B) oscillates about the fixed-point value (C) at $x_{+}(\sigma,q)$.}
\label{fig:xz_2}
\end{figure}

We are now ready to derive an expression for the canard critical parameter $\sigma_{c}(q,\delta) = \sigma_{0}(q) + \delta\,\sigma_{1}(q) + \cdots$ derived from the invariant manifold $\sigma\,z \equiv Z = \Phi(x,q,\delta)$ in the limit $\delta \ll 1$. As noted in Fig.~\ref{fig:Oregon_fixed}, a canard explosion is expected to occur near the maximum $x_{B}(q)$, as seen in Fig.~\ref{fig:xz_1}, while a canard implosion is expected to occur near the minimum $x_{D}(q)$, as seen in Fig.~\ref{fig:xz_2}.

Using the Oregonator-2 equations \eqref{eq:x2_prime}-\eqref{eq:z2_prime}, we obtain the canard perturbation equation
\begin{equation}
\delta\,\left( \sigma\,x \;-\frac{}{} \Phi\right) \;=\; \pd{\Phi}{x}\;\left[ x\,(1 - x) - \left(\frac{x - q}{x + q}\right)\;\Phi \right],
 \label{eq:O2_canard}
\end{equation}
where, by defining the function $\psi(x,q) = (x-q)/(x+q)$, we have the partial derivatives evaluated at $\epsilon = 0$:
\begin{equation}
\left. \begin{array}{rcl}
F_{Z0} &=& -\,\psi(x,q) \\
F_{\sigma0} &=& 0 \\
G_{Z0} &=& -\,1\\ 
G_{\sigma0} &=& x
\end{array} \right\},
\end{equation}
and thus $h(x) = 0$ in Eqs.~\eqref{eq:canard_eq_1} and \eqref{eq:canard_S1}. Here, the lowest-order $x$-nullcline function 
\begin{equation} 
\Phi_{0}(x,q) \;=\; x\,(1 - x)/\psi(x,q) \;=\; x\,(1-x)\,\left(\frac{x+q}{x-q}\right)
\label{eq:Phi0_O2}
\end{equation}
has a maximum $x_{B}(q)$ and a minimum 
$x_{D}(q)$, which are the two positive roots \eqref{eq:x1_min}-\eqref{eq:x2_max} of the cubic polynomial \eqref{eq:cubic_O2}. At each one of those two roots, we can write the factorization
\begin{equation}
\Phi_{0}^{\prime}(x,q) \;=\; (x - x_{0})\;\Psi_{0}(x,q), 
\end{equation}
where $x_{0}(q) = x_{B}(q)$ or $x_{D}(q)$, and
\begin{equation}
\Psi_{0}(x,q) \;=\; -\;[2\,x^{2} \;+\; A_{0}(q)\,x \;+\; B_{0}(q)]/(x - q)^{2},
\end{equation}
with $A_{0}(q) = 2\,x_{0}(q) - (1 + 2q)$ and $B_{0}(q) = 2\,x_{0}^{2}(q) - (1+2q)\,x_{0}(q) + 2q\,(1-q)$.

At first order in $\delta$, Eq.~\eqref{eq:O2_canard} yields
\begin{equation}
\sigma_{0}\,x \;-\; \Phi_{0}(x) \;=\; -\,(x - x_{0})\;\Psi_{0}(x)\,\psi(x)\;\Phi_{1}(x).
\label{eq:F1_eq}
\end{equation}
Since we want $\Phi_{1}(x)$ to be finite at $x = x_{0}$, we now require 
\begin{equation}
\sigma_{0}(q) \;=\; \Phi_{0}(x_{0},q)/x_{0} \;=\; (1-x_{0})\,\left(\frac{x_{0} + q}{x_{0} - q}\right),
\label{eq:sigma_0}
\end{equation}
which yields $\sigma_{B}(q)$ at $x_{0} = x_{B}$ and $\sigma_{D}(q)$ at $x_{0} = x_{D}$. With Eq.~\eqref{eq:sigma_0}, we now use the factorization
\[ \sigma_{0}(q)\,x \;-\; \Phi_{0}(x,q) \;=\; (x - x_{0})\,H_{1}(x,q), \]
where
\begin{equation} 
H_{1}(x,q) \;=\; \frac{x\;[x\,(x_{0} - q) - q\,(x_{0} +q - 2)]}{(x_{0} - q)\,(x - q)},
 \end{equation}
so that Eq.~\eqref{eq:F1_eq} yields 
\[ \Phi_{1}(x) \;=\; K_{1}(x) \;\equiv\; -\,H_{1}(x)/[\psi(x)\Psi_{0}(x)], \]
where
\begin{equation}
K_{1}(x,q) \;=\; x\left(\frac{x+q}{x_{0}-q}\right) \frac{P_{1}(x,q)}{Q_{0}(x,q)},
\end{equation}
with $P_{1}(x,q) = (x_{0}-q)\,x - q\,(x_{0}+q-2)$ and $Q_{0}(x,q) = 2\,x^{2} \;+\; A_{0}(q)\,x \;+\; B_{0}(q)$.
 
At second order in $\delta$, we can now calculate the functions $R_{2}(x)$ and $S_{1}(x)$ from Eqs.~\eqref{eq:canard_R2}-\eqref{eq:canard_S1}:
\begin{equation}
\left. \begin{array}{rcl} 
R_{2}(x) &=& K_{1}(x) \left[ -\,\psi(x)\;K_{1}^{\prime}(x) \;+\frac{}{} 1 \right] \\
 && \\
 S_{1}(x) &=& x
 \end{array} \right\},
 \end{equation}
so that $\sigma_{1}(q)$ is now defined at the critical fixed-point $x_{0}(q)$ according to Eq.~\eqref{eq:canard_a1}:
\begin{equation}
\sigma_{1}(q) \;=\; \frac{K_{1}(x_{0})}{x_{0}} \left[ -\,\psi(x_{0})\;K_{1}^{\prime}(x_{0}) \;+\frac{}{} 1 \right].
\label{eq:sigma_1}
\end{equation}
Using the model-value for $q = 7.62 \times 10^{-5}$, we first look at the canard explosion near the maximum $x_{0}(q) = x_{B}(q)$, where $\sigma_{0B}(q) = 0.500229$ and $\sigma_{1B}(q) = 0.250305$, so that for $\delta = 0.001$, we find 
\begin{equation}
\sigma_{cB}(q,\delta) \;=\; \sigma_{0B}(q) \;+\; \sigma_{1B}(q)\,\delta \;=\; 0.500479, 
\end{equation}
which is in excellent agreement with the numerical value $0.500479$ shown in Fig.~\ref{fig:xz_1}. Next, we look at the canard implosion near the minimum $x_{0}(q) = x_{D}(q)$, where $\sigma_{0D}(q) = 2.41346$ and $\sigma_{1D}(q) = -\,2.73923$, so that for $\delta = 0.001$, we find 
\begin{equation}
\sigma_{cD}(q,\delta) \;=\; \sigma_{0D}(q) \;+\; \sigma_{1D}(q)\,\delta \;=\; 2.41072, 
\end{equation}
which is again in excellent agreement with the numerical value $2.4106$ shown in Fig.~\ref{fig:xz_2}. 

\begin{figure}
\epsfysize=1.8in
\epsfbox{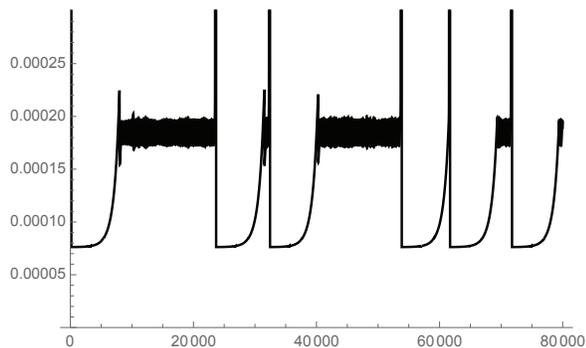}
\caption{Mixed-mode oscillations for $a = 2.4107$ near the critical parameter value.}
\label{fig:fig_23}
\end{figure}

Lastly, Fig.~\ref{fig:fig_23} shows that, as we approach the canard implosion, the Oregonator-2 solutions enter into a regime of mixed-mode oscillations (MMO) in which large-amplitude relaxation oscillations alternate with small-amplitude oscillations about the steady-state solution \cite{Popovic_2008,Guckenheimer_2011,Desroches_2012}.

\subsection{Validity of the Oregonator-2 model}

\begin{figure}
\epsfysize=1.7in
\epsfbox{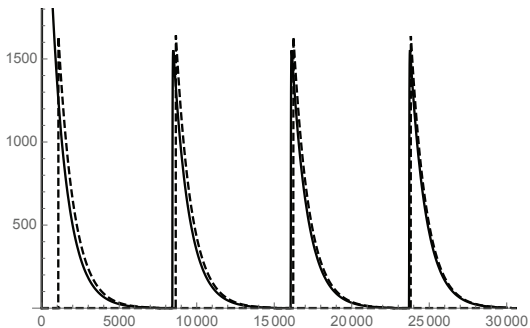}
\caption{Numerical Oregonator-3 solutions (solid) and numerical Oregonator-2 solutions (dashed)  for $\sigma = 0.55$: $y_{3}(t)$ (solid) compared to the constraint solution $y_{2}(t) = \sigma z_{2}(t)/(q + x_{2}(t))$ (dashed) obtained from the numerical Oregonator-2 solutions $x_{2}(t)$ and $z_{2}(t)$.}
\label{fig:xyz_32}
\end{figure}

We conclude this Section by discussing the numerical evidence in support of the validity of the reduction from the Oregonator-3 equations \eqref{eq:x_prime}-\eqref{eq:z_prime} to the Oregonator-2 equations \eqref{eq:x2_prime}-\eqref{eq:z2_prime}. First, in Fig.~\ref{fig:xyz_32}, we see that the Oregonator-3 numerical solution $y_{3}(t)$ and the constraint equation $y_{2}(t) = \sigma\,z_{2}(t)/[q + x_{2}(t)]$, constructed from the Oregonator-2 numerical solutions $x_{2}(t)$ and $z_{2}(t)$, show very good agreement, with comparable amplitudes and periods, as can be seen in Fig.~\ref{fig:Oregonator_2_Period}. Second, we note that the phase-space portrait seen in Fig.~\ref{fig:vdP_period} is nearly indistinguishable when constructed with the Oregonator-3 solutions $z_{3}(t)$ versus $x_{3}(t)$ and the Oregonator-2 solutions $z_{2}(t)$ versus $x_{2}(t)$. Lastly, the Oregonator-3 canard explosion shown in Fig.~\ref{fig:xyz_3D} occurs at a value of the critical parameter that is identical to the critical parameter seen in the Oregonator-2 canard explosion shown in Fig.~\ref{fig:xz_1}.

\section{Conclusions}

In this paper, we have performed the asymptotic analysis of the limit-cycle period of relaxation oscillations and the critical parameters for the canard explosion and implosion associated with the large-amplitude relaxation-oscillation solutions of two important examples of oscillating chemical reactions. For both the CIMA reactions (Sec.~\ref{sec:CIMA}) and the Oregonator model of the BZ reactions (Sec.~\ref{sec:Oregonator}), the asymptotic limit of the relaxation-oscillation periods \eqref{eq:P_th} and \eqref{eq:O_2_period}, respectively, show excellent agreements with numerical periods, as seen in Figs.~\ref{fig:P_th} and \ref{fig:Oregonator_2_Period}, respectively. 

In addition, using the Fenichel perturbation analysis \cite{Fenichel_1979}, the perturbative calculations of the critical parameter for the canard explosion in the CIMA model and the canard explosion and implosion in the Oregonator-2 model have shown excellent agreements with numerical values.

\acknowledgments

The work presented in this paper was funded by the NSF grant DUE-1742241.

\appendix

\section{\label{sec:cubic}Roots of a Cubic Polynomial}

In this paper, we have on several occasions needed to find the roots of a cubic polynomial with coefficients that may depend on a set of model parameters. Here, we consider the generic cubic polynomial
\begin{equation}
P(x) \;=\; 4\,x^{3} \;+\; a\,x^{2} \;+\; b\,x \;+\; c,
\label{eq:cubic_P}
\end{equation}
where $(a,b,c)$ are real-valued coefficients. The three roots $x_{i}(a,b,c)$ ($i = 1,2,3$) are explicitly calculated from the trigonometric identity 
\begin{equation}
4\,\beta^{3}\cos^{3}(\phi/3) \;-\; 3\,\beta^{3}\cos(\phi/3) \;-\; \beta^{3}\cos\phi \;=\; 0,
\label{eq:trig_id}
\end{equation}
where the amplitude $\beta$ and the phase $\phi$ may be real or complex valued. 

\begin{figure}
\epsfysize=1.8in
\epsfbox{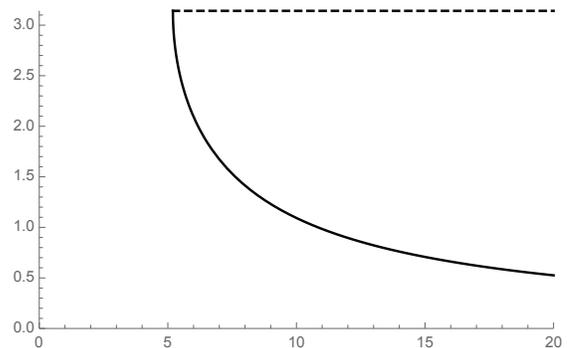}
\caption{Plot of $\phi(a)$ versus $a$ in the range $\sqrt{27} \leq a \leq 20$, where the dashed line corresponds to $\phi(\sqrt{27}) = \pi$.}
\label{fig:fig_24}
\end{figure}

In order to find the first cubic root of Eq.~\eqref{eq:cubic_P}, we first remove the quadratic term by inserting the translation $x = \alpha + z$: $P(\alpha + z) = 4\,z^{3} + \frac{1}{2}\,P^{\prime\prime}(\alpha)\,z^{2} + P^{\prime}(\alpha)\,z + P(\alpha)$ and require that $P^{\prime\prime}(\alpha) = 24\,\alpha + 2\,a = 0$, i.e., $\alpha = -\,a/12$. From the trigonometric identity \eqref{eq:trig_id}, therefore, we find the root $z_{1} = \beta\,\cos(\phi/3)$, where
\begin{eqnarray*}
P^{\prime}(\alpha) &=& 12\,\alpha^{2} + 2a\,\alpha + b \;=\; -\,12\,\alpha^{2} + b \;\equiv\; -\,3\,\beta^{2}, \\
P(\alpha) &=& 4\,\alpha^{3} + a\,\alpha^{2} + b\,\alpha + c \;=\; -\,8\,\alpha^{3} + b\,\alpha + c \\
 &\equiv& -\,\beta^{3}\,\cos\phi,
\end{eqnarray*}
so that $\beta(a,b) = (a^{2}/36 - b/3)^{\frac{1}{2}}$ is either real or imaginary, and $\cos\phi(a,b,c) = \gamma(a,b,c) \equiv (-\,a^{3}/27 + b\,a/12 - c)/\beta^{3}$. Here, if $\beta$ is real, then the phase $\phi$ is either real (i.e., $0 \leq \phi \leq \pi$) if $-1 \leq \gamma \leq 1$, or the phase $\phi = i\,\psi$ is imaginary if $\gamma \geq 1$, or the phase $\phi = \pi - i\,\psi$ is complex if $\gamma \leq -1$, where $\cosh\psi = |\gamma|$. If $\beta = i\,|\beta|$ is imaginary, on the other hand, then $\phi = \pi/2 - i\,\psi$ is complex, with $\sinh\psi = (-\,a^{3}/27 + b\,a/12 - c)/|\beta|^{3}$. The other two $z$-roots are easily found to be $z_{2,3} = -\,\beta\,\cos(\pi/3 \pm \phi/3)$, so that the three roots of Eq.~\eqref{eq:cubic_P} are
\begin{eqnarray*}
x_{1}(a,b,c) & = & -\,a/12 \;+\; \beta(a,b)\,\cos[\phi(a,b,c)/3], \\
x_{2}(a,b,c) & = & -\,a/12 \;-\; \beta(a,b)\,\cos[\pi/3 + \phi(a,b,c)/3], \\
x_{3}(a,b,c) & = & -\,a/12 \;-\; \beta(a,b)\,\cos[\pi/3 - \phi(a,b,c)/3].
\end{eqnarray*}
When the phase $\phi$ is real, the roots are labeled so that $x_{3} \leq x_{2} \leq x_{1}$, and the roots $x_{1} = x_{2}$ and $x_{2} = x_{3}$ merge when $\phi = \pi$ and $\phi = 0$, respectively

For example, consider the CIMA cubic polynomial $2 x^{3} + a\,(1 - x^{2})$ found in Eq.~\eqref{eq:cubic_CIMA}, from which we obtain 
\begin{equation}
\left. \begin{array}{rcl}
(\alpha,\; \beta) &=& (a/6,\; a/3) \\
 && \\
 \gamma &=& 1 - 54/a^{2} \;=\; \cos\phi(a)
 \end{array} \right\}. 
 \end{equation}
 Hence, if $a \geq \sqrt{27}$, all three roots are real since the phase $0 < \phi(a) \leq \pi$ (see Fig.~\ref{fig:fig_24}), including one negative root and two positive roots $x_{E} < 0 < x_{D} < x_{B}$: 
 \begin{eqnarray*}
x_{1}(a) & = & x_{B}(a) \;=\; a/6 + (a/3)\,\cos[\phi(a)/3], \\
x_{2}(a) & = & x_{D}(a) \;=\; a/6 - (a/3)\,\cos[\pi/3 + \phi(a)/3], \\
x_{3}(a) & = & x_{E}(a) \;=\; a/6 - (a/3)\,\cos[\pi/3 - \phi(a)/3].
\end{eqnarray*}
For $0 < a < \sqrt{27}$,  on the other hand, we find $\gamma < -1$ and $\phi = \pi - i\,\psi$, where $\psi(a) = {\rm arccosh}(54/a^{2} - 1)$, so that $x_{1} = x_{2}^{*} = a/6 + (a/3)\,\cos(\pi/3 - i\,\psi/3)$, i.e., the minimum and maximum of the $x$-nullcline for the CIMA model have merged and become complex-valued, and $x_{3} = a/6 - (a/3)\,\cosh(\psi/3)$.


\begin{thebibliography}{99}

\bibitem{Noyes_Field_1974} R.~M.~Noyes and R.~J.~Field, Annu.~Rev.~Phys.~Chem.~{\bf 25}, 95 (1974).

\bibitem{Zhabotinsky_1991} A.~M.~Zhabotinsky, Chaos {\bf 1}, 379 (1991).

\bibitem{Cervellati_Greco_2017} R. Cervellati and E. Greco, J.~Chem.~Educ.~{\bf 94}, 195 (2017).

\bibitem{FKN_1972} R.~J.~Field, E. K\"{o}r\"{o}s, and R.~M.~Noyes, J. Am. Chem. Soc.~{\bf 94}, 8649 (1972). 

\bibitem{Winfree_1972} A. T. Winfree, Science {\bf 175}, 634 (1972). 

\bibitem{Clarke_1974} B. L. Clarke, J. Chem. Phys. {\bf 60},1481 \& 1493 (1974). 

\bibitem{Strogatz_2015} S.~H.~Strogatz, {\it Nonlinear Dynamics and Chaos}, 2nd ed. (Westview Press, 2015).

\bibitem{Diener_1984} M.~Diener, Math.~Intell.~{\bf 6}, 38 (1984).

\bibitem{Wechselberger_2007} M. Wechselberger, {\it Canards}, Scholarpedia {\bf 2}, 1356 (2007).

\bibitem{Krupa_2001} M. Krupa and P. Szmolyan, J.~Diff.~Eqs.~{\bf 174}, 312 (2001).

\bibitem{Fenichel_1979} N.~Fenichel, J.~Diff.~Eqs.~{\bf 31}, 53 (1979).

\bibitem{Ginoux_2011} J.-M.~Ginoux and J.~Llibre, J.~Phys.~A: Math.~Theor.~{\bf 44}, 465203 (2011).

\bibitem{Footnote_VdP} In its standard dimensionless form \cite{Strogatz_2015}, the Van der Pol equation is expressed in terms of the time coordinate normalized to the frequency $\omega$, which means that Eq.~\eqref{eq:vdp_eq} is expressed as $x^{\prime\prime} - \varepsilon\,(1 - x^{2})\, x^{\prime} + x = a$, where $\varepsilon \equiv \nu/\omega \gg 1$. Hence, our small parameter $\epsilon \equiv \varepsilon^{-2} \ll 1$.

\bibitem{Algaba_2020} A. Algaba, K.-W. Chung, B.-W. Qin, and A.J. Rodr\'{i}guez-Luis, Physica D {\bf 406}, 132384 (2020).

\bibitem{Lengyel_1990} I. Lengyel, G. R\'{a}bai, and I.~R. Epstein, J. Am. Chem. Soc.~{\bf 112}, 9104 (1990).

\bibitem{Lengyel_1991} I. Lengyel and I.~R. Epstein, Science {\bf 251}, 650 (1991).

\bibitem{Epstein_1995} I.~R. Epstein and I. Lengyel, Physica D {\bf 84}, 1 (1995).

\bibitem{Awal_Epstein_2020} N.M. Awal and I.~R. Epstein, Phys.~Rev.~E {\bf 101}, 042222 (2020).

\bibitem{Field_Noyes_1974} R.~J.~Field and R.~M.~Noyes, J.~Chem.~Phys.~{\bf 60}, 1877 (1974).

\bibitem{Tyson_Fife_1980} J.~J.~Tyson and P.~C.~Fife, J.~Chem.~Phys.~{\bf 73}, 2224 (1980), see Table I.

\bibitem{Brons_1991} M. Br\o ns and K. Bar-Eli, J. Phys. Chem. {\bf 95}, 8706 (1991).

\bibitem{Field_2007} R.~J.~Field, {\it Oregonator}, Scholarpedia {\bf 2}, 1386 (2007).

\bibitem{Szmolyan_2001} P. Szmolyan and M. Wechselberger, J.~Diff.~Eqs.~{\bf 177}, 419 (2001).

\bibitem{Brons_footnote} For example, by using a different normalization $(\ov{x},\ov{y},\ov{z},\ov{\tau}) = (x/q,y,z/q,q\tau/\epsilon)$, Br\o ns and Bar-Eli \cite{Brons_1991} obtain a constraint equation for $\ov{x}(\ov{y})$ by setting $(q/\epsilon)\,d\ov{x}/d\ov{\tau} = \ov{y} + \ov{x}\,(1 - \ov{y}) - q\,\ov{x}^{2} = 0$ and, thus, obtain a two-dimensional model expressed in terms of the independent dimensionless variables $(\ov{y},\ov{z})$.

\bibitem{Popovic_2008} N. Popovi\'{c}, J.~Phys.: Conf.~Ser.~{\bf 138}, 012020 (2008).

\bibitem{Guckenheimer_2011} J. Guckenheimer and C. Scheper, SIAM J. Appl. Dyn. Syst.~{\bf 10}, 92 (2011).

\bibitem{Desroches_2012} M. Desroches, J. Guckenheimer, B. Krauskopf, C. Kuehn, H. M. Osinga, M. Wechselberger, SIAM Rev.~{\bf 54}, 211 (2012).


\end{thebibliography}
\end{document}